\documentclass[aps,prx,reprint,floatfix,amsmath,amssymb,
superscriptaddress, nofootinbib, keywords]{revtex4-2}

\usepackage{amsfonts}
\usepackage{amsmath}
\usepackage{amssymb}
\usepackage{graphicx}
\usepackage{epstopdf}
\usepackage{color}
\usepackage{hyperref}
\usepackage{xcolor}

\usepackage{bm}
\usepackage{printlen}
\usepackage{units}
\usepackage{enumerate}
\usepackage{comment}

\renewcommand{\vec}[1]{\mathbf{#1}}
\newcommand{\Torder}{\text{T}_{\tau} }
\newcommand{\Trace}{\operatorname{Tr}}

\renewcommand{\Re}{\text{Re}}

\newcommand{\Ham}{{\mathcal H}}

\usepackage{booktabs}
 
\hypersetup{hidelinks}

\begin{document}

\title{
Unconventional superfluidity and quantum geometry of  topological bosons
}

\author{Ilya Lukin}
 \affiliation{V. N. Karazin Kharkiv National University, Svobody Sq. 4, 61022 Kharkiv, Ukraine}
 \affiliation{Kharkiv Institute of Physics and Technology, Akademichna 1, 61108 Kharkiv, Ukraine}  
  
\author{Andrii Sotnikov}
 \affiliation{V. N. Karazin Kharkiv National University, Svobody Sq. 4, 61022 Kharkiv, Ukraine}
 \affiliation{Kharkiv Institute of Physics and Technology, Akademichna 1, 61108 Kharkiv, Ukraine}

\author{Alexander Kruchkov}

\affiliation{Institute of Physics, {\'E}cole Polytechnique F{\'e}d{\'e}rale de Lausanne,  Lausanne, CH 1015, Switzerland}   
  
\affiliation{Department of Physics, Harvard University, Cambridge, Massachusetts 02138, USA}  

\affiliation{Branco Weiss Society in Science, ETH Zurich, Zurich, CH 8092, Switzerland}

\begin{abstract} 
We investigate superfluidity of bosons in gapped topological bands and discover a new phase that has no counterparts in the previous literature. This phase is characterized by a highly unconventional modulation of the order parameter, breaking the crystallographic symmetry,  and for which the condensation momentum is neither zero
nor any other high-symmetry vector of the Brillouin zone. 
This unconventional structure impacts the spectrum of Bogoliubov excitations and,  consequently,  the speed of sound in the system.  Even in the case of perfectly flat bands,  the speed of sound and Bogoliubov excitations remain nonvanishing,  provided that the underlying topology and quantum geometry are nontrivial.   Furthermore,  we derive detailed expressions for the superfluid weight using the Popov hydrodynamic formalism for superfluidity and provide estimates for the Berezinskii–Kosterlitz–Thouless temperature,  which is significantly enhanced by the nontriviality of the underlying quantum metric. These results are applicable to generic topological bosonic bands,  with or without dispersion.  To illustrate our findings,  we employ the Haldane model with a tunable bandwidth,  including the narrow lowest-band case. 
Within this model, we also observe a re-entrant superfluid behavior: As the Haldane's magnetic flux is varied, the Berezinskii–Kosterlitz–Thouless transition temperature initially decreases to almost zero, only to resurface with renewed vigor.
\end{abstract}

\maketitle

Quantum geometry is a fundamental concept that governs the distance and entanglement between quantum states in the Hilbert space \cite{Provost1980,Ma2010}. Although extensively studied in the field of quantum information, particularly through the  Fischer information metric and related Fubini-Study metric \cite{Anandan1990, Braunstein1994, Facchi2010}, its application in condensed matter has been relatively overlooked.
The analysis of the quantum geometry of electronic states by means of the Fubini-Study metric provides with a valuable information about the system's topological properties \cite{Marzari1997}, flatness of electronic bands \cite{Kruchkov2022, Wang2021}, electric polarization \cite{Resta1999,Souza2000,Resta2011}, and anomalous transport observables \cite{Neupert2013, Jackson2015, Peotta2015, Rhim2020,  Kozii2021, Rossi2021, Ahn2022, Kruchkov2023, Kaplan2023}.  
While the quantum transport framework from quantum-geometric perspective has been established for the fermionic systems \cite{Neupert2013, Peotta2015,  Kruchkov2023}, its investigation in the bosonic counterparts is still relatively limited. This is primarily due to the intricate complexities associated with the bosonic behavior at low temperatures.

Bosons at low temperatures form a quantum-coherent state known as the Bose-Einstein condensate (BEC) below a certain critical temperature $T_C$ (typically in the range of hundreds of nanoKelvin for laser-cooled atomic gases). The energy-momentum dispersion relation of bosons plays a crucial role in determining the fundamental properties of the condensate, including the critical temperature $T_C$ and the nature of elementary excitations within the interacting condensate. For instance, in the case of a quadratic bare dispersion $\varepsilon({\vec k}) = {\hbar^2 k^2}/{2 m}$, the critical temperature for Bose-Einstein condensation in three dimensions can be expressed as
\begin{align}
k_B T_C \sim \frac{\hbar ^2 n^{2/3}}{m}. 
\label{TC}
\end{align}
Moreover, the renowned Bogoliubov result \cite{Bogoliubov1947} relates the speed of sound in the BEC to the local interaction strength $U$, condensate density $n_0$, and the mass $m$ of the boson through the expression:
\begin{align}
c = \left( {\frac{U n_0}{m}} \right)^{1/2}.
\label{SoundSpeed}
\end{align}
This Bogoliubov result holds qualitatively even when the dispersion deviates from the quadratic form, as long as the boson mass is well-defined. However, the case of dispersionless quantum states, i.e., bosonic states with an almost constant dispersion $\varepsilon({\vec k})\simeq\text{const}$ (also known as a flat band), challenges the conventional concepts of Bose-Einstein condensation \cite{Huber2010,You2012,Tovmasyan2013,Julku2016,Julku2021}. 

In the presence of a flat band, where the effective mass diverges ($m\to\infty$), the quantities given in Eqs.~\eqref{TC} and \eqref{SoundSpeed} loose their well-defined meanings. Generally, superfluidity is not expected when the particles become infinitely heavy, i.e., when the bandwidth is of orders of magnitude smaller than the interaction strength. 
In contrast to conventional condensates, where all the condensed particles occupy the lowest energy state (e.g., with $\vec k_* =0$ in the case of quadratic dispersion), the presence of a perfectly flat band presents an obstacle to the formation of a Bose-Einstein condensate. In this scenario, bosonic states experience frustration and have \textit{no preferred momentum} $\vec k_*$ corresponding to the lowest noninteracting state. As a result, the conventional concept of homogeneous BEC is no longer applicable, and the condensation can occur at any \textit{finite} momentum $\vec k_* \ne 0$ \cite{Abrikosov1975}. In such cases, the underlying topology of ultracold bosons becomes crucial for determining their behavior.

In particular, the \textit{gapless} flat bands on the kagome lattice~\cite{Julku2021,Julku2022} give rise to a qualitatively different Bogoliubov BEC excitations with the speed of sound
\begin{align}
c  \sim  U n_0 \sqrt{ \text{Re} [\mathfrak G_{ii} (\vec k_*)] } ,
\label{SoundSpeed2}
\end{align}
where $\vec k_*$ is the condensation wave vector and  $\mathfrak G_{ii} (\vec k)$ is the quantum-geometric tensor \cite{Provost1980}, defined on the flat band Bloch states $| u_{\vec k} \rangle$ as 
\begin{align}
\mathfrak{G}_{ij} (\vec k) = \langle \partial_i u_{\vec k} |  
\left( 
1- {|  u_{\vec k} \rangle   \langle  u_{\vec k} | }
\right)
 |  \partial_j u_{\vec k} \rangle,
\label{metrics}
\end{align}
\noindent
where $\partial_i \equiv \frac{\partial}{\partial k_i}$. The imaginary part  of $\mathfrak{G}$ is responsible for topology, and determines the (off-diagonal) Berry curvature $ \mathcal F_{ij} = \text{Im} \mathfrak{G}_{ij}$; the real part $ \mathcal G_{ij} = \text{Re} \mathfrak{G}_{ij}$ is the Fubini-Study metrics and is responsible for the band geometry and its flatness \cite{Kruchkov2022}. 
It is important to note that the expression~\eqref{SoundSpeed2} is applicable specifically to the kagome lattice and it cannot be straightforwardly generalized to establish a simple relationship between the bosonic observables and the quantum metrics of flat bands in a generic setting.

This study delves into the broader realm of Bose-Einstein condensation and superfluidity by examining a more general scenario involving \textit{gapped} flat topological bands, as outlined in Ref.~\cite{Kruchkov2022}. Our specific focus revolves around the exploration of gapped flat Chern bands, which can be realized in various systems, including the representative case of the Haldane model \cite{Haldane1988}. In this model, one or both energy bands may exhibit a  flat structure \cite{Neupert2011, Kruchkov2022}.
While we generally observe the possibility of Bose-Einstein condensation and superfluidity in flat bands, with a certain limit linked to the underlying quantum geometry \cite{Julku2021}, it is advisable to approach the system with caution. It is important to note that the explicit influence of quantum metrics appears in a simplified scenario, and a more comprehensive treatment involves a more intricate tensor structure, which we elucidate further in this paper. In the subsequent sections, we derive: (i) the superfluid weight; (ii) estimates for the Berezinskii-Kosterlitz-Thouless (BKT) transition temperature $T_{\text{BKT}}$; (iii) Bogoliubov excitations; and (iv) the speed of sound in flat band Bose-Einstein condensates.
In contrast to previous studies that relied on the effective Bogoliubov theory \cite{Julku2021, Julku2022}, we employ the vigorous Popov hydrodynamic theory of superfluidity. This theory, known for its absence of infrared singularities, allows us to delve into the profound exploration of the superfluid properties exhibited by topological bosons.  Furthermore, since achieving a perfectly flat topological band on local Hamiltonians is hindered by Wannier obstructions \cite{Kruchkov2022, Chen2014}, it is necessary to consider the scenario of nearly flat bands instead. Our results apply to dispersive bands with nontrivial topology, including both the narrow-band and dispersive cases. By contrasting these two, we reveal fundamental distinctions in the phase diagrams of superfluidity, the order parameter structure, characterized by the different impact of quantum geometry and interactions.

Our key findings can be summarized as follows:
(i) We identify three distinct Bose-Einstein condensate phases in topological nearly-flat bands: BEC-i, BEC-ii, and the novel BEC-iii (see Table~\ref{tbl:2} and the phase diagram in Fig.~\ref{phasediagram}).
(ii) While BEC-i has similarities to conventional condensation in dilute atomic gases and BEC-ii shows resemblance to a previous study~\cite{Julku2021}, the newly discovered BEC-iii phase has no analogs in the existing literature on Bose gases.
(iii) The order parameter structure in the BEC-iii phase is highly nontrivial (see Fig.~\ref{Fig3}), affecting the spectrum of Bogoliubov excitations and, consequently, the speed of sound in the system. Remarkably, the speed of sound in this phase, which scales as $\sqrt{U n_0}$, is enforced by the nontriviality of the underlying quantum metrics.
(iv) The BEC-iii phase exhibits pronounced superfluidity even when the bandwidth is minimized, indicating the impact of quantum metrics of the underlying Bloch states. However, obtaining a simple analytical expression for the superfluid weight in terms of quantum metrics and lower bounds, as in the fermionic analog~\cite{Peotta2015, Rossi2021}, remains challenging due to the higher complexity of the problem.
(v) Re-entrant superfluidity: In the narrow-band regime of the Haldane model, the superfluid behavior decreases as the Haldane's flux grows (phase BEC-i), reaches nearly zero, and then re-emerges with the increased strength at higher Haldane fluxes. The re-entrant superfluidity, characterized by re-entrant BKT temperature, is depicted as a function of the Haldane's magnetic flux (see Fig.~\ref{Haldane-flux}).
Overall, our results uncover rich features in the quantum  behavior and superfluid properties of the topological bosons, highlighting the intricate interplay between topology, quantum metrics, bandwidth, and interactions.

The aforementioned findings hold particular significance for experiments involving ultracold gases in optical lattices, where notable advancements have been made in engineering Haldane Hamiltonians and manipulating their parameters \cite{Jotzu2014}.  However, directly measuring Fubini-Study metrics presents a considerable experimental challenge and is currently feasible only for a limited number of systems \cite{Yu2020}. Notably, measuring the speed of sound in Bose-Einstein condensates serves as a highly accurate experimental tool \cite{Andrews1997, Bohlen2020}, offering potential insights into the underlying quantum geometry of nontrivial bosonic states, and can be implemented to probe condensed system with nontrivial topology.

\section{ Bose-Einstein condensation in nearly-flat bands in the mean-field approximation}\label{sec:2}

\subsection{ Noninteracting Hamiltonian: Which flat bands are good for BEC? }

Not all flat bands support Bose-Einstein condensation. Consider a trivial flat band belonging to the ``atomic insulator'' class \cite{Kruchkov2022}, such as a 1D chain with a lattice parameter $a_0$ approaching infinity ($a_0 \to \infty$). Despite featuring a perfectly flat band with $\varepsilon=0$, this flat band is trivial. The individual wave functions are widely separated, making the formation of a coherent BEC state unattainable. Altering the system's dimensionality or symmetry does not alleviate this issue as long as the bosonic wave functions remain strongly localized on the lattice sites.
To facilitate the BEC state, one needs the underlying Bloch state to be topological.\footnote{
More precisely, as we will explore in detail later, the underlying Bloch states must possess a non-trivial quantum-geometric tensor.}

\begin{figure*}.  \includegraphics[width=0.95 \textwidth]{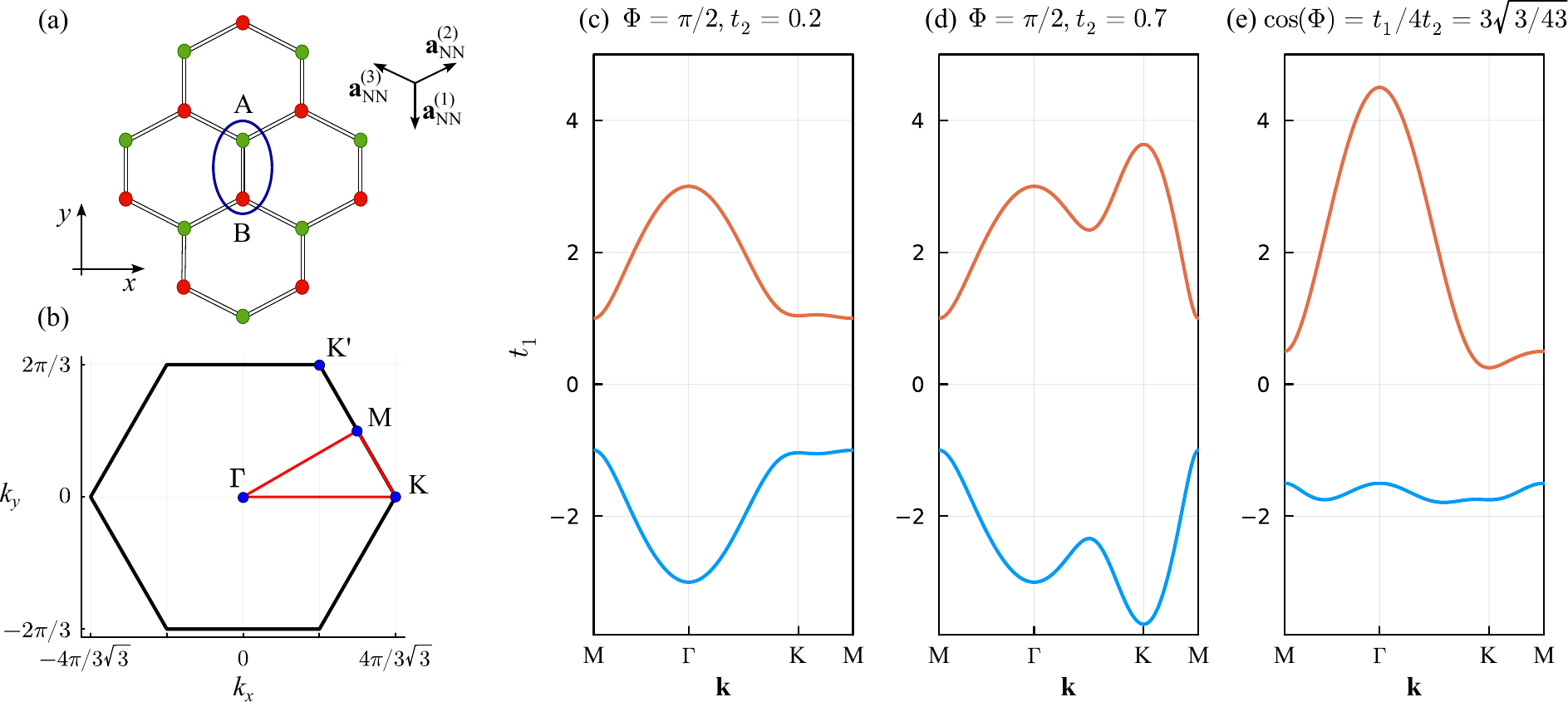} 
  \caption{\label{Fig1}%
Formation of flat bands in the Haldane model:
(a) The real-space lattice structure and the two-site unit cell in the Haldane model.
(b) Brillouin zone with the path connecting the high-symmetry points M, K, and $\Gamma$.
(c)-(e) Noninteracting band structure at different parameters $\Phi$ and $t_2$.
   }
\end{figure*}

Looking at the problem from a different perspective, we consider a lattice described by a precise tight-binding model that incorporates all the hopping terms $t_{ij}$,
\begin{align}
     \Ham = \sum_{ ij } t_{ij} b^{\dag}_{i} b^{}_{j}  ,
\label{Ham1}
\end{align}
where $b_i$ and $b_i^\dagger$ are the bosonic annihilation and creation operators, respectively, acting on the lattice site~$i$.
In practice, the complex-valued hopping amplitudes~$t_{ij}$ can extend beyond the nearest-neighbor (NN) and next-nearest-neighbor (NNN) hopping processes. While a perfectly flat topological band would require infinite hopping range \cite{Kruchkov2022}, in reality, the hopping amplitudes decay exponentially with increasing distance. Therefore, we can truncate the Hamiltonian \eqref{Ham1} to include only the first few nearest neighbors:
\begin{align}
    \Ham = & t_{\text{NN} } \sum_{\langle ij \rangle} b^{\dagger}_{i} b^{}_{j}  + t_{\text{NNN}}  \sum_{\langle \langle ij \rangle \rangle}   b^{\dagger}_{i} b^{}_{j}
+ ... \, ,
\end{align}
where $\langle ij \rangle$ and $\langle \langle ij \rangle \rangle$
denote the NN and NNN  
bonds, respectively.  

In the minimal setup to create a topological band, we require at least two bonds, with one of the hopping terms being complex-valued:
\begin{align}
    \mathcal{H} = t_1 \sum_{\langle ij \rangle} b_i^\dagger b_j + t_2 \sum_{\langle\langle ij \rangle\rangle} e^{i \Phi} b_i^\dagger b_j + \text{h.c.},
\label{Ham0}
\end{align}
This truncated Hamiltonian is known as the Haldane model \cite{Haldane1988}. Here, $t_1$ and $t_2$ are real-valued parameters in energy units, and $\Phi$ represents the Haldane flux (modulo $2 \pi$). The band structure of this Hamiltonian, determined by the choices of $t_1$, $t_2$, and $\Phi$, consists of two topological bands. These bands are typically not flat (see also Fig.~\ref{Fig1}).
The Fourier-transformed Hamiltonian $\mathcal{H}^{\alpha \beta}(\mathbf{k})$, where $\alpha, \beta = A,B$ are sublattice indices, in the two-orbital basis is given by (see also Refs.~\cite{Haldane1988,Neupert2011})
\begin{align}
    \Ham^{\alpha\beta}(\vec k) = \vec h (\vec k) \cdot \boldsymbol \sigma + h_0 (\vec k)  \sigma_{0},
  \label{haldane1}
\end{align}
where $\boldsymbol \sigma$$=$$(\sigma_{x},\sigma_{y},\sigma_{z})$ are the Pauli matrices,   $h_{0}(\vec k)$$=$$2t_{2}\cos{\Phi} \sum_{i=1}^{3} \cos{\vec k  \, \vec a^{(i)}_{\rm NNN}}$, $\vec h$$=$$(h_x, h_y, h_z)$, 
$h_{x}(\vec k)$$=$$t_{1} \sum_{i=1}^{3} \cos{\vec k  \,\vec a^{(i)}_{\rm NN}}$, 
    $h_{y}(\vec k)$$=$$t_{1} \sum_{i=1}^{3} \sin{\vec k \, \vec a^{(i)}_{\rm NN}}$, 
$h_{z}(\vec k)$$=$$-2t_{2}\sin{\Phi}\sum_{i=1}^{3} \sin{\vec k  \, \vec a^{(i)}_{\rm NNN}}$. Here     
$\vec a^{(i)}_{\rm NN}$  and $\vec a^{(i)}_{\rm NNN}$  are coordinate vectors connecting NN and NNN atomic sites. 
We shall use
$\vec a^{(1)}_{\rm NN}$$=$$(0, -1) a_0$, $\vec a^{(2)}_{\rm NN}$$=$$( \frac{\sqrt{3}}{2}, \frac{1}{2}) a_0$, $\vec a^{(3)}_{\rm NN}$$=$$(-\frac{\sqrt{3}}{2}, \frac{1}{2})a_0$ for the nearest neighbors [see also Fig.~\ref{Fig1}(a)], and 
$\vec a^{(1)}_{\rm NNN}$$=$$(\sqrt{3}, 0) a_0$, $\vec a^{(2)}_{\rm NNN}$$=$$(-\frac{\sqrt{3}}{2}, \frac{3}{2})$, $\vec a^{(3)}_{\rm NNN}$$=$$(-\frac{\sqrt{3}}{2}, -\frac{3}{2}) a_0$ for the next-nearest neighbors; here $a_0$ is the closest distance between the two neighboring atoms. In what follows, we set $a_0=1$, $\hbar=1$, and $t_1 =1$.

The  band structure of the Haldane model \eqref{haldane1} depends on the relative strength of $t_1$ and $t_2$. We illustrate this in Fig.~\ref{Fig1}(c)--(e).  
The bandwidth can be carefully controlled by tuning the hopping strength $t_2/t_1$ and flux $\Phi$. 
One can see that both the bandwidth and the band gap can be tuned by varying $t_2/t_1$, and the lower band (LB) becomes very narrow. It is important to note that when using the NNN truncation, the resulting Chern band cannot be perfectly flat \cite{Kruchkov2022}, regardless of the values chosen for $t_1$ and $t_2$. However, by optimizing the parameters of the Hamiltonian~\eqref{Ham0}, it is possible to obtain a topological band with a significantly reduced bandwidth/bandgap ratio, $w/\Delta \ll 1$.  

We focus on the effective Haldane model \eqref{haldane1} to derive the central results of this paper. Although we consider the  Haldane model on the honeycomb lattice with a two-site unit cell, the conclusions hold true for other lattices and symmetries that preserve topological bands. Our findings extend to generic topological bands with different topological invariants, emphasizing the importance of the underlying structure of the (orbital-independent) quantum-geometric tensor.

\subsection{Interacting Hamiltonian}
Let us consider the minimal interacting Hamiltonian characterized by the on-site interaction strength $U$. In the second quantized form, it can be written as  
\begin{align}
    \Ham_{\rm tot} = \Ham + U \sum_{i} n_{i}(n_{i} -1) - \mu \sum_{i} n_{i}.
\label{Ham-int}
\end{align}
Here $\Ham$ represents the non-interacting Hamiltonian associated with a nearly-flat topological band [see, e.g., Eq.~\eqref{Ham0}], while $n_i=b^{\dag}_i b^{}_i$ denotes the on-site density. The chemical potential $\mu$ regulates the average particle density. In the scenario of condensation within the flat band, it tends towards the lowest-available energy state in the  \textit{interacting} model, which may not necessarily correspond to $\vec k_*=0$. Owing to the underlying translational symmetry of the lattice, the condensate in the Bloch bands is assumed to be described by a fixed wave vector ${\vec k}_{*}$ along with two complex-valued parameters $\xi_{A}$ and $\xi_{B}$, representing densities and phases on the sublattices $\alpha=A,B$ [refer to Fig.~\ref{Fig1}(a)]. These parameters can be further rescaled as follows:
\begin{align}
 \xi_{\alpha} \to  \sqrt {n_{0}}   \xi_{\alpha},
 \quad 
 \sum_{\alpha}  |\xi_{\alpha}|^{2}  = 1, \label{eq:condit}
\end{align}
Here, $n_{0}$ denotes the condensate density within the unit cell. Consequently, the problem of determining the Bose-Einstein condensate wave function can be reformulated as the constrained minimization of the following functional:
\begin{equation}\label{Energy}
    \mathcal E (\vec k) = \xi ^{\dagger} \mathcal H(\vec k) \xi + U n_{0} (|\xi_{A}|^{4} + |\xi_{B}|^{4}), 
\end{equation}
where ${\cal H}(\vec k)$ represents the Fourier transform of the non-interacting Hamiltonian with a flat or dispersive topological band~\eqref{Ham0} and $\xi=[\xi_A,\xi_B]^T$. To facilitate our analysis, we express Eq.~\eqref{Energy} as ${\cal E}(\vec k)={\cal K}(\vec k) + {\cal U}(\vec k)$, where ${\cal K}(\vec k)$ corresponds to the kinetic part (dispersive contribution to the total energy), and ${\cal U}(\vec k)$ represents the interaction part. It is important to note that the minimization is performed over both $\xi$ and $\vec k_*$.
As a result, we find the \textit{condensate wave vector} $\vec k_*$, which minimizes the energy ${\cal E}(\vec k)$. By definition, the derivative vanishes at the condensate wave vector $\vec k_*$,
\begin{align}
\left . \partial_{\vec k} {\cal E}(\vec k_{}) \right|_{\vec k_*} =  \left. \xi^{\dagger} \partial_{\vec k} \mathcal H(\vec k) \xi  \right|_{\vec k_*} = 0.
\label{minimum}
\end{align}
The expression~\eqref{minimum} is useful and will be employed below in the evaluation of the superfluid weight.

When translational symmetry undergoes spontaneous breaking, the assumption of the condensate wave function characterized by a single specific value of $\vec k_{*}$ is no longer applicable. Instead, we employ a more general ansatz by considering a linear combination of multiple contributions $\xi_{\alpha j} \exp(-i\vec k^{(j)}_* \vec r)$, involving different condensation wave vectors $\vec k^{(j)}_*$. The mean-field energy, determined within this wave-function ansatz, is then optimized over all amplitudes  $\xi_{\alpha j}$ and specific wave vectors~$\vec k^{j}_*$. Further details regarding this optimization process will be presented in subsequent sections.

\textbf{Minimization procedure}. To determine the global minimum of $\mathcal E (\vec k)$ with respect to the variational parameters $\vec k$ and $\xi_\alpha$, subjected to the constraint~\eqref{eq:condit}, we employ a two-step minimization approach. \textit{In the first step,} we fix the wave vector $\vec k$ and utilize the imaginary-time evolution to find the extremum parameters $\xi_\alpha$ for that specific $\vec k$.
\textit{In the second step,}  we calculate $\mathcal E (\vec k)$ for different $\vec k$ values and determine the minimum. It is important to note that the second step corresponds to the mean-field approximation of the interacting (renormalized) band structure.

We emphasize the significance of the first step. In models with a \textit{perfectly} flat lowest band, both terms in Eq.~\eqref{Energy} can be simultaneously minimized by selecting $\xi$ from the flat band and subsequently finding $\vec k$ that results in a uniform distribution of $\xi$ within the unit cell. However, in truncated models with the finite-range hopping given by Eq.~\eqref{Ham0}, the lowest band is not perfectly flat. As a result, there is a competition between the kinetic and interaction energies, and contributions from upper bands can become significant. This competition can lead to a nonuniform density distribution within the unit cell and a marginal ``dressing'' of the condensate due to the effects of higher excited bands.

\subsection{BEC phases in nearly flat bands and phase transitions between them}

\begin{table}[t]
\caption{Summary of different Bose-Einstein-condensed phases in narrow topological bands.}\label{tbl:2}
\centering
\begin{tabular}{@{}l l  l @{}}
\toprule
Phase  & Feature  &  Cause\\
 \midrule
BEC-i &  Condensation at $\Gamma$ point & small $t_2/t_1$  \\
BEC-ii & Condensation at K point  & moderate $t_2/t_1$ \\
BEC-iii  &  Condensation inside the BZ & intermediate $t_2/t_1$ \\
 \bottomrule
\end{tabular}
\end{table}

In this section, we examine various condensate phases that arise from minimizing the interacting mean-field energy \eqref{Energy}. The interplay between interactions, topology, and the narrow bandwidth gives rise to three distinct phases, which we refer to as BEC-i, BEC-ii, and BEC-iii (see also Table~\ref{tbl:2}).
Before delving into the main results concerning condensation in the topological bands, let us examine the lowest-band dispersion relation and the structure of the interaction density in the lowest band, $|\xi_{A}|^{4} + |\xi_{B}|^{4}$, for different values of the model parameters $t_{2}$ and $\Phi$ (all energy-related quantities are measured in units of $t_1=1$). Analytically, it can be shown that the minimum of the interaction density is determined by the equation:
\begin{align}
    \sin{\Phi} \sum_{i=1}^3 \sin{\left[ \vec k  \cdot \vec a^{(i)}_{\rm NNN} \right]} = 0.
\label{min-int-dens}
\end{align}
For our choice of coordinate system [see notations below Eq.~\eqref{haldane1}], the  solution of Eq. \eqref{min-int-dens} for the nontrivial Haldane flux ($\Phi\neq0$), is defined by $k_{x}=0$ plus $C_3$ rotations.

Our analysis reveals three distinct qualitative scenarios arising from the interplay between the kinetic and interacting terms. The first two scenarios occur at high symmetry points of the Brillouin zone, while the third scenario exhibits a more elegant behavior:

\textbf{BEC-i: Condensation at $\Gamma$ point}: The noninteracting term ${\cal K}(\vec k)$ favors condensation at the $\Gamma$ point ($\vec k_* = 0$), which also minimizes the interaction energy density ${\cal U}(\vec k)$. This scenario is \textit{stabilized by both interactions and dispersion}, and the associated Bogoliubov excitations depend on the residual dispersion. This phase finds an analogy with the conventional BEC in ultracold gases.

\textbf{BEC-ii: Condensation at {\rm K} points}:
Energy minimization results in the condensation at the K (K$'$) points. 
In a wide range of model parameters 
the mean-field transition from BEC-ii to BEC-i can be driven by the change of $t_2$ and $\Phi$ entering the noninteracting model rather than by the change of $Un_0$.
The latter dependence can appear in the system under study by going beyond the mean-field approximation.

\textbf{BEC-iii: Condensate wave vector anywhere in the Brillouin zone}: 
The interacting term ${\cal U}(\vec k)$ drives the condensation momentum \textit{away from the high-symmetry points} associated with the minima of non-interacting term ${\cal K}(\vec k)$.  
 In this case, a variation in the interaction strength can  impact the position of the condensate wave vector $\vec k_*$ and also result into the ``dressing'' effects mediated by the upper bands. 
 Notably, this phase is not confined to high-symmetry points, unlike the BEC-i and BEC-ii scenarios.

\subsubsection{Phase diagram and the structure of order parameters.}

\begin{figure*}[t]
    \centering
    \includegraphics[width=\textwidth]{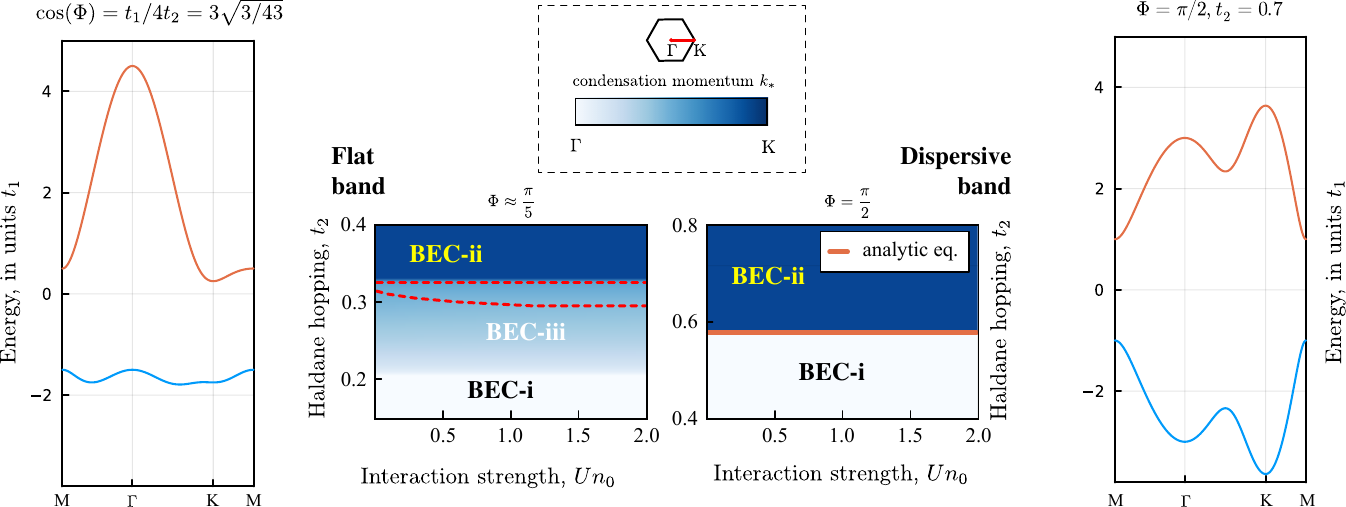}
    \caption{
    Noninterecting band structures and phase diagrams of the BEC phases for the cases of nearly-flat (left) and dispersive (right) bands. 
    The diagrams are obtained in the mean-field approximation and plotted in the parameter space of $t_2$ (Haldane hopping strength) and $U n_0$ (interaction strength), given in dimensionless units normalized by $t_1$.
    The conventional phase (BEC-i) is characterized by condensation at the $\Gamma$ point of the Brillouin zone (white).
    BEC-ii is characterized by the condensation at the K (K$'$) point of the Brillouin zone (blue).
    BEC-iii is characterized by a wavelength modulation incommensurate with the underlying lattice, allowing for $\vec k_*$ within the Brillouin zone (gradients of blue).
    The phase boundary for the ``dispersive band'' case indicated with the red solid line can be determined by Eq.~\eqref{phase boundary}, while for nearly-flat bands the red dashed line denotes modification to BEC-iii$'$, as discussed in Eq. \eqref{BEC-iv} and below. 
    }
    \label{phasediagram}
\end{figure*}

This section explores the phase diagram of the Haldane model \eqref{haldane1}, focusing on two distinct scenarios of nearly flat and dispersive topological bands. The main findings are summarized in Fig.~\ref{phasediagram}. These phase diagrams are derived using the mean-field approximation and represent the parameter space of $t_2$ (Haldane hopping strength) and $U n_0$ (interaction strength), with dimensionless units normalized by $t_1$.
In the following discussion, we delve into the properties of each phase, with a particular focus on the novel phase BEC-iii.

\textit{BEC-i phase and its properties.} 
The BEC-i phase has been extensively studied in previous literature. It corresponds to a uniform Bose-Einstein condensate, where the condensation occurs at the conventional point $\vec k_* = 0$, which corresponds to the $\Gamma$ point of the Brillouin zone. 
This phase remains stable even in the absence of Haldane hoppings ($t_{2} = 0$). In this limit, the Haldane model reduces to the Bose-Hubbard model, and the BEC-i phase is adiabatically connected to the trivial BEC in the Hubbard model.
At the $\Gamma$ point, the wave function of the lowest band is homogeneous across the sublattices, minimizing the interaction energy. Additionally, the contribution from the dispersion relation to the energy, which is determined by the lowest-band wave function, is also minimized. Consequently, the condensate is entirely composed of states from the lowest band.
In Sec.~\ref{sec:3}, we demonstrate that the excitations in this phase exhibit rather trivial properties. The quantum-geometric contributions from the upper bands vanish at the $\Gamma$ point. Therefore, the BEC-i phase can be regarded as a conventional condensate phase, and for further properties, we refer to relevant textbooks on the subject.

\textit{BEC-ii phase and its properties.}
The BEC-ii phase, which has been partially studied in the previous literature (for the gapless case of kagome lattice) \cite{Julku2021,Julku2022}, is characterized by a real-space modulation, where the condensate structure is commensurate with the lattice. For the \textit{gapped} topological bands, this phase generally emerges from the BEC-i phase due to the competition between different minima in the noninteracting band structure, as shown in Fig.~\ref{Fig1}(d). In the regime of a moderate Haldane hopping $t_{2}\sim t_1$, the dispersion relation can exhibit a minimum at the K(K$'$)-point of the Brillouin zone (BZ). The position of the condensate momentum $\vec{k}_*$ is not sensitive to the interaction strength, assuming reasonable values of $Un_0$.
Within a conventional mean-field ansatz, the lowest-band wave function at the K(K$'$)-point is concentrated entirely on a single sublattice, either $A$ or $B$. The interaction energy for this density configuration is maximized and significantly surpasses the interaction energy at the $\Gamma$-point for the same total density. Consequently, at the K-point, the interaction energy and the dispersive contribution compete with each other. However, our numerical analysis using a trial wave function ansatz indicates that the condensate wave function solely resides in the lowest band, similar to the BEC-i phase.

In the upcoming section, we present a Bogoliubov excitation analysis that reveals potential instabilities in the trial wave function for the BEC-ii phase, which exhibits translational invariance. These instabilities occur at the wave vector $\vec{K'}$, while the condensate wave vector is located at $\vec{K}$.
The structure of the Bogoliubov instabilities suggests a modification to the wave function ansatz. 
The BEC-ii has degenerate dispersive energies $\cal K$ at the points $\vec K$ and $\vec K'$,  while the interaction energy $\cal U$ becomes maximal as soon as the condensate is formed solely at one of these points.
This degeneracy implies that a wave function of the form $\xi_{\alpha \vec{K}} \exp(-i \vec{K} \cdot \vec{r}) + \xi_{\alpha \vec{K'}} \exp(-i\vec{K'} \cdot \vec{r})$ will possess the \textit{same noninteracting energy} $\mathcal{K}(\vec{K})$ yet \textit{lower interaction energy} $\mathcal{U}$. Indeed, our numerical calculations with the modified trial wave function
\begin{align}
 \psi_\alpha = \xi_{\alpha \Gamma} + \xi_{\alpha  \vec K} \exp(-i\vec K \vec r) + \xi_{\alpha \vec K'} \exp(-i\vec K' \vec r)
\label{trial wf}
\end{align}
confirm this statement. The wave function in real space can be expressed as
\begin{equation}
    | {\rm{BEC\text{-}ii} } \rangle = \sqrt{\frac{n_{0}}{2}} \left[
    {\exp(-i \vec K \vec r)}, {\exp(-i \vec{K'} \vec r)}
    \right]^T,
    \label{trial wf 2}
\end{equation}
where the vector coordinates correspond to the sublattices $A$ and $B$. Written in this form, the condensate components with different wave vectors are concentrated on different sublattices.
The wave function \eqref{trial wf 2} minimizes the mean-field energy by simultaneously minimizing both the non-interacting and interacting components.
First, since the density is uniformly distributed across the lattice, the sum $\sum_{\alpha} |\psi_{\alpha}|^{4}$ approaches its minimum value of $1/2$.
Second, the wave function can be expressed as $| \text{BEC-ii} \rangle = \sqrt{n_0/2} [|u_{0 \vec{K}}\rangle \exp(-i\vec{K} \cdot \vec{r}) + |u_{0 \vec{K'}}\rangle \exp(-i\vec{K'} \cdot \vec{r})]$, utilizing the properties of Bloch functions for the wave vectors $\vec{K}$ and $\vec{K'}$ on sublattices $A$ and $B$.
Substituting this into the expression for the dispersive energy, we obtain $\frac{n_0}{2} [ \varepsilon_{0}(\vec{K}) + \varepsilon(\vec{K'})] = n_0 \varepsilon (\vec{K})$, which minimizes the non-interacting energy as long as $\vec{K}$ corresponds to the minimum position of the lowest Bloch band $|u_{0 \vec{k}}\rangle$.
Consequently, the modified phase \eqref{trial wf 2} is generally more stable in the regime of large $Un_{0}$ compared to the previous ansatz of Ref.~\cite{Julku2021} (without $\vec K'$ contributions to the order parameter); Eq. \eqref{trial wf 2} represents the mean-field  ground state  for the \textit{gapped topological bands}. This modification exhibits additional unique properties.

\textit{BEC-iii phase and its properties.}  The BEC-iii phase, to the best of our knowledge, has not been explored in previous literature. 
To elucidate the origin of this phase, let us consider a \textit{gedanken experiment} in which the lowest Haldane band becomes perfectly flat through  engineering the infinite-range hoppings \cite{Kruchkov2022}. In this situation, the noninteracting bosons do not possess discernible energy minima, since their dispersion relation is featureless. As a result, neither the $\Gamma$ point nor the K point are favored. This introduces frustration to the behavior of condensing particles, allowing them to select any condensation wave vector within the Brillouin zone upon the inclusion of interactions.

\begin{figure}
  \includegraphics[width=\linewidth]{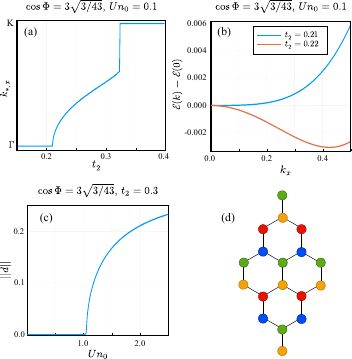} 
  \caption{\label{Fig3}%
   Characteristics of different BEC phases in the model under study: (a) the dependence of the $\vec k_{*}$ (its $x$-component) on $t_{2}$ for other parameters fixed; (b) the MF energies $\mathcal {E} (\vec k)$ across the transition between BEC-iii and BEC-i phases, which show the continuous nature of the transition; (c) the norm of $d_{\alpha}$ component of the wave function in the BEC-iii$'$ phase; (d) the real-space modulation of the condensate density in the BEC-iii$'$ phase coded by four different colors.  }
\end{figure}

In practice, achieving perfectly flat Haldane bands is hindered by the Wannier obstructions \cite{Kruchkov2022}. However, the case of nearly-flat bands still captures the intricate interplay between mean-field terms that lead to condensation at momenta $\vec{k}_*$ located outside the high-symmetry points. Our numerical calculations (see Fig.~\ref{Fig3}) indicate that the BEC-iii phase emerges in the vicinity of the parameters corresponding to the nearly-flat bands, which in the Haldane model are determined by the relationship between the hopping $t_{2}$ and the Haldane flux $\Phi$.
Nevertheless, the symmetry of the underlying Hamiltonian dictates the condensate wave vectors $\vec{k}_*$ to position on the symmetry lines (in our chosen coordinate frame, these lines are given by $k_{y} = 0$ and its $C_3$ rotations). An example of the dispersion relation leading to this phase is depicted in Fig. \ref{Fig3}(a), where the minimum of the lowest-band dispersion occurs at the intermediate point between K and $\Gamma$.

The competition between interacting and noninteracting terms in the BEC-iii phase is intense. As a result, the condensate wave vector  is determined by the strength of interactions, $\vec{k}_* = \vec{k}_*(Un_{0})$, leading to condensation vectors beyond the local minima of the noninteracting model. The influence of the dispersive contribution becomes significant as the bandwidth increases, making the condensate order highly dependent on the microscopic details of the system (parametrized by $t_{2}/t_1$ and phase $\Phi$ in our case). A more detailed analysis reveals that the condensate wave function may also exhibit certain \textit{interband effects}, which are determined by the microscopic characteristics of the system. Consequently, the BEC-iii phase demonstrates a high level of tuning and a rich variety of real-space order parameters, contingent upon the underlying microscopic details.

The order parameter of the BEC-iii phase exhibits remarkable complexity. A straightforward trial wave function with a single modulation $\vec k_*$ proves to be potentially unstable when examining Bogoliubov excitations. Specifically, the primary instability arises at the wave vector $\vec k_{*} + [0, 2\pi/3]$ (we have  the interatomic distance $a_0=1$). Through a detailed analysis of the Bogoliubov instabilities, we propose the following conjecture for the condensate wave function in the BEC-iii phase:\footnote{
Note that the trial wave functions based on the superpositions of the rotated states with $(C_3)^n \vec k_*$ and/or $(C_6)^n \vec k_*$ exhibit higher energy compared to the wave function \eqref{BEC-iv}. }
\begin{equation}\label{BEC-iv}
    \psi_{\alpha \vec k_{*}  } = \xi_{\alpha} \exp[-i\vec k_{*} \vec r] + d_{\alpha} \exp[-i (\vec k_{*} + \vec b/2) \vec r], 
\end{equation}
where ${\vec b}  = [0, 4\pi/3]$ is a reciprocal lattice basis vector.

\textit{Flat-band BEC-iii'}. Numerical calculations unequivocally demonstrate the second-order transition at moderate $U n_{0}$ that distinguishes regions with zero and nonzero values of $d_{\alpha}$, as it is shown in Fig.~\ref{Fig3}(c). This transition signifies the emergence of the anomalous order parameter, leading to the formation of the BEC-iii$'$ phase. 
It is worth noting that the order parameter \eqref{BEC-iv} with the unconventional modulation BEC-iii$'$ arises in the region characterized by both the flattest band and the strongest effect of interactions (indicated by the dashed red line in Fig.~\ref{phasediagram}).
Notably, the BEC-iii$'$ phase expands the unit cell, since its order parameter is not an eigenvector of lattice translations. This happens because the flat-band scenario with strong interactions brings significant frustration to the condensation process. Nonetheless, the modified BEC-iii$'$ phase retains similarities to the generic BEC-iii phase, particularly in terms of the real-space modulation representing the frustrated nature of bosonic condensation in the flat band.

\subsubsection{Phase transitions}

The two phase diagrams, reflecting two cases of dispersive band and narrow band at chosen parametrizations of Haldane model,  are shown  in Fig.~\ref{phasediagram}.  These diagrams provide insights into various types of phase transitions:
(i) In  dispersive bands, a direct phase transition occurs from the BEC-i phase to the BEC-ii phase, without an intermediate BEC-iii phase.
(ii) In narrow bands, additional phase transitions emerge: from the BEC-i phase to the BEC-iii phase, and from the BEC-iii phase to the BEC-ii phase.
(iii) At high $U n_{0}$, there is a possibility of an additional phase transition.

Within the mean-field (MF) approximation, we performed additional analysis confirming that the transition from the BEC-ii phase to the BEC-i phase, without an intermediate BEC-iii phase, is of the first-order nature. The dispersion energy exhibits minima at the $\Gamma$ and K points of the Brillouin zone, while the interaction energy remains the same in both phases.
Since the condensate wave function resides entirely within the lowest band for both phases, we can precisely compute the MF energies at the $\Gamma$ and K points. This allows us to determine the boundary between these phases by comparing the corresponding MF energies: $\varepsilon_{0}(0) + \frac{U n_{0}}{2} = \varepsilon_{0}(\vec {K}) + \frac{U n_{0}}{2}$, where we evaluate the sums of $|\xi_{A}|^{4}$ and $|\xi_{B}|^{4}$ in both phases. This leads to the following relation governing the phase boundary:
\begin{equation}
3 t_{2} \cos{\Phi} + \sqrt{3} t_{2} \sin{\Phi} - t_{1} = 0.
\label{phase boundary}
\end{equation}
In Fig.~\ref{phasediagram}, we have depicted this MF boundary for the transition between the BEC-i and BEC-ii phases, which is further corroborated by numerical analysis.

In the presence of the intermediate BEC-iii phase, we observe distinct phase transitions characterized by the dependence of the condensate wave vector $\vec k_{*}$ (specifically its $x$ component) on the parameter $t_{2}$ while keeping $\Phi$ and $U n_{0}$ fixed, as illustrated in Fig.~\ref{Fig3}(a). Notably, the system undergoes multiple phase transitions.
The transition from the BEC-iii to BEC-i phase at small $t_{2}$ is continuous, evident from the smooth change in MF energy across the phase boundary depicted in Fig.~\ref{Fig3}(b). The minimum of the mean-field energy continuously shifts from the $\Gamma$ point to a nonzero $\vec k_{*}$.
Similarly, the transition between the BEC-iii and BEC-iii$'$ phases is also of the second order. This is evident from Fig.~\ref{Fig3}(c), where the behavior of the norm of the additional wave function component $d_{\alpha}$ [defined in Eq.\eqref{BEC-iv}] is illustrated.

\section{Quantum Geometry and Bogoliubov Excitations}\label{sec:3}

In the preceding section, we employed the mean-field approximation to analyze Bose-Einstein condensation in both dispersive and nearly flat bands. Our investigation revealed the existence of three distinct mean-field ground states: BEC-i, BEC-ii, and BEC-iii. Now, we proceed by incorporating quantum fluctuations and examine their properties within these identified phases, with particular emphasis on the new superfluid, BEC-iii.

\subsection{Effective Hamiltonian}

We can express the Hamiltonian \eqref{Ham-int} in the momentum space  by introducing  creation and annihilation operators $b^{\dagger}_{\vec k \alpha}$ and $b^{}_{\vec k \alpha}$,  where $\alpha$ represents the sublattice index,
\begin{align}
  \Ham_{\rm tot}  = \sum_{a, b} \sum_{\vec k} b^{\dagger}_{\vec k \alpha} 
  \left[\Ham^{\alpha\beta}(\vec k) - \mu \delta_{\alpha\beta} \right] b^{}_{\vec k  \beta}  \nonumber
    \\
    +   \frac{U}{N} \sum_{\alpha} \sum_{\vec k,\vec k',\vec q}  b^{\dagger}_{\vec k+\vec q\alpha} b^{\dagger}_{\vec k'-\vec q\alpha} b^{}_{\vec k\alpha}  b^{}_{\vec k'\alpha} .
\end{align}
The Fourier-transformed Hamiltonian of the flat band \eqref{Ham0} is denoted as $\Ham^{\alpha\beta}(\vec k)$, $\mu$ is the chemical potential, and $N$ is the number of unit cells. In the Bogoliubov approach, the chemical potential can be determined through the equation: 
\begin{align}\label{chemical}
    \mu 
    =  \sum_{\alpha, \beta} \overline{\xi}_{\alpha } \mathcal H^{\alpha\beta}(\vec k_{*}) \xi_{ \beta} + 2 U n_{0} \Theta . 
\end{align}
Here, $\xi_{\alpha}$ represents the mean-field solutions (complex numbers), and $\Theta$ corresponds to the expectation value of the Gross-Pitaevskii-type  nonlinearity,
\begin{align}
  \Theta =   \sum_{\alpha} |\xi_{\alpha}|^{4}.
 \label{Theta} 
\end{align}
By replacing the condensate wave functions with the expectation values of the mean-field solution \eqref{Energy},
\begin{equation}
b_{\vec k_{*}\alpha} \to \sqrt{N n_{0}} \xi_{\alpha},
\end{equation}
we can neglect higher-order terms and focus on quadratic terms involving the remaining operators. 
The reduced Hamiltonian contains terms representing the annihilation of excitations into the condensate, given by $b_{\vec k_* -\vec q} b_{\vec k_* + \vec q}$. Additionally, it introduces mixing between operators with wave vectors $(\vec k_* +\vec q)$ and $(\vec k_* - \vec q)$.
The resulting \textit{Bogoliubov Hamiltonian}  $\mathbb{H}$ can be expressed as a block matrix with twice the dimension compared to the original Hamiltonian:
\begin{align}
    \mathbb{H}(\vec k_{*} + \vec q) = 
\begin{pmatrix}
  H (\vec q)   & 2 U n_{0} \xi_{\alpha}^{2} \delta_{\alpha\beta} \\
    2 U n_{0} \overline{\xi_{\alpha}}^{2} \delta_{\alpha\beta} &     H^* (-\vec q)  
\end{pmatrix}   .
\label{BHam}
\end{align}
Here, the diagonal elements are determined by
\begin{align}\label{Hq}
    H (\vec q) =   \Ham^{\alpha\beta}(\vec k_{*} + \vec q) - \mu \delta_{\alpha\beta} + 4 U n_{0} |\xi_{\alpha}|^{2} \delta_{\alpha\beta} .
\end{align}
The Bogoliubov Hamiltonian \eqref{BHam} acts on the eigenstates of the form $[b^{}_{\vec k_{*}+\vec q\alpha}; b^{\dagger}_{\vec k_{*}-\vec q\alpha}]^{T}$.

In the trivial   phase (BEC-i), the mean-field solution promotes uniform values $|\xi_{\alpha}|^{2} = 1/\mathcal A$, where $\mathcal A$$=$$2 a_0^2$ represents the unit cell size in real space (we set $a_0=1$ further on). Consequently, the terms $4 U n_{0} |\xi_{\alpha}|^{2} \delta_{\alpha\beta}$  can be  absorbed into the renormalization of the chemical potential $\mu$. In contrast, the convenience of this operation is limited to the BEC-i phase, while the spatially modulated phases (in particular, BEC-iii) lack such a straightforward simplification.  
As we shall uncover, the presence of this term intricately modifies the dispersion relation of the Bogoliubov excitations in the original BEC-ii phase without the modified wave function ansatz and BEC-iii.

To determine the Bogoliubov modes, we diagonalize the matrix $\tilde{\mathbb{H}} = \sigma_{z} \mathbb{H}$ numerically, considering various condensate wave vectors and interaction strengths.
Our numerical calculations consistently reveal the presence of gapless modes with a linear dispersion around $\vec q=0$ (\textit{Goldstone modes}), while further Bogoliubov excitations exhibit a finite gap in the studied cases. In the  subsequent material, we provide the analytical expressions for the Bogoliubov modes and the speed of sound,  rigorously compare our analytical findings with the precise numerical calculations, and elucidate the role of quantum geometry.

\subsection{Analytical solution for the Bogoliubov modes}  

\begin{figure}[t]
 \includegraphics[width=\linewidth]{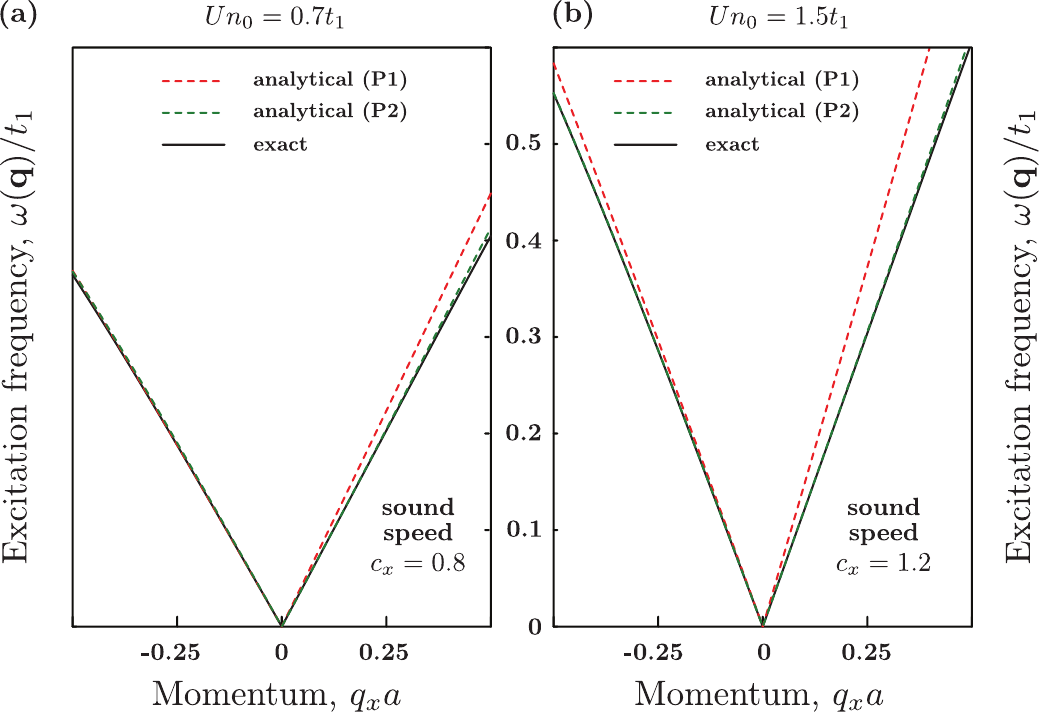} 
 \caption{\label{fig:Bogoliubov_Dispersions}%
 Excitation spectrum of the Bogoliubov modes at small $|q| = |\vec k_* - \vec k|$ in the BEC-iii phase ($\Phi=\pi/5, t_{2} = 0.3$),  for two different values of  $U n_{0}$. The dispersion relations are computed with both the exact Bogoliubov Hamiltonian $\mathbb{H}$ - $\omega_{ex}$ and with different projected Hamiltonians $\mathbb{H}_{P}$ - $\omega_{P1}$ (with the lowest band projection) and $\omega_{P2}$ (with the projection on $|\xi_{ \vec k_{*} + \vec q}\rangle$). 
   }
\end{figure}

To obtain analytical results, we project the Bogoliubov Hamiltonian onto the lower band. Mathematically, this is carried by replacing the former operators  $[b_{\vec k_{*}+\vec q\alpha}; b^{\dagger}_{\vec k_{*}-\vec q\alpha}]^{T}$ with the new operators $[u_{0,\vec k_{*}+\vec q\alpha} b_{0}; \overline{u}_{0,\vec k_{*}-\vec q\alpha} b^{\dagger}_{0}]^{T}$. The physical meaning of  $u_{0,\vec k_{*}+\vec q\alpha}$ is the lower-band Bloch wave function with the wave vector $\vec k_{} + \vec q$ on the site $\alpha$, and $b^{\dag}_{0}$ is the bosonic creation operator in the lower band. 
The \textit{projected Bogoliubov Hamiltonian} $\mathbb{H}_{\text{P}}$ reads as 
\begin{equation}
    \mathbb{H}_{\text{P}} (\vec k_{*} + \vec q)   = 
    \begin{pmatrix}
    \epsilon( \vec q)     & 2 U n_{0} \, \eta (\vec q)  \\
        2 U n_{0} \, \eta^* (\vec q)  &  \epsilon( - \vec q) 
    \end{pmatrix} ,
\end{equation}
where 
\begin{align}
  \epsilon( \vec q)  =  \varepsilon_0(\vec k_{*}+\vec q) - \mu + 4 U n_{0} \sum_{\alpha} |\xi_{\alpha}|^{2} |u_{\vec k_{*}+\vec q\alpha}|^{2} . 
\end{align}
Here $\varepsilon_0(\vec q)$ represents the dispersion relation of the lower band prior to condensation, and 
\begin{align}
    \eta (\vec q) =  \sum_{\alpha} \xi_{\alpha}^{2} \overline{u}_{\vec k_{*}-\vec q\alpha} \overline{u}_{\vec k_{*}+ \vec q\alpha}  . 
\end{align} 
Upon the projection, the Bogoliubov Hamiltonian may no longer remain gapless since $\xi_{\alpha}$ and $u_{0,\vec k_{}\alpha}$ may differ. This  is typically resolved by self-consistently adjusting the chemical potential beyond the mean-field expectation in Eq.~\eqref{chemical}, thereby restoring gaplessness. Alternatively, one can access the analytical solution by neglecting the higher-band corrections, which are typically of the order of $Un_{0}/\Delta$ ($\Delta$ is the single-particle gap). In this case, the condensate wave function $\xi_{\alpha}$ remains exclusively within the lower band. Under this additional assumption, the Hamiltonian becomes exactly gapless under the conditions:
\begin{align}
 \xi_{\alpha} = u_{\vec k_{*}\alpha} , \quad      
 \mu = \varepsilon_{0}(\vec k_{*})+2 U n_{0} \Theta.
\end{align}
(For compactness, in this section we use the notation for the lower band $u_{0,\vec k \alpha} \equiv u_{\vec k \alpha}$). Numerical verification confirms that the projected Hamiltonian yields Goldstone modes consistent with the full Hamiltonian within a reasonable range of the interaction strength $U n_{0}$ (see Fig.~\ref{fig:Bogoliubov_Dispersions}).

To find the spectrum of the Bogoliubov excitations and the speed of sound, we expand the projected Bogoliubov Hamiltonian $\mathbb{H}_{\text{P}}$  up to the second order in $\vec q$, 
\begin{align}
 \label{expansion} 
  \mathbb{H}_{\text{P}} (\vec q)
    = 
    2 U n_{0} \Theta (\sigma_0 + \sigma_x)  + \mathcal V _i q_i \sigma_z + \mathcal W_{ij} q^i q^j \sigma_0   &  \nonumber 
   \\
   +   \overline{\mathcal Y_{ij}} q^{i} q^{j} \sigma^{-} +  {\mathcal Y_{ij}} q^{i} q^{j} \sigma^{+}   &. 
\end{align}
 where $\sigma_i$ are the conventional spin-1/2 Pauli matrices, $i=\{x,y,z\}$, $\sigma^{\pm} = (\sigma_x \pm i \sigma_y)/2$, and $\sigma_0$ is the unit matrix.  Here, $\mathcal V_{i}$ has the physical meaning of the renormalized single-particle velocity $ c_{i} =\partial_{i}  \varepsilon_{0}(\vec k )$,   
\begin{align}
    \mathcal V_{i} = &    4 U n_{0} \sum_{\alpha} |u_{\vec k_{*}\alpha}|^{2}(u_{\vec k_{*}\alpha} \partial_{i} \overline{u}_{\vec k_{*}\alpha} + \overline{u}_{\vec k_{*}\alpha} \partial_{i} u_{\vec k_{*}\alpha}) \nonumber 
    \\
    + &  \partial_{i} \varepsilon_{0}(\vec k_{*})  ,
    \nonumber 
\end{align}
 $\mathcal W_{ij}$ has the physical meaning of the renormalized  inverse effective mass  $ m^{-1}_{ij} = \partial_{i} \partial_{j} \varepsilon_{0}(\vec k)$, 
\begin{align}
    \mathcal  W_{ij} = &  2 U n_{0} \sum_{\alpha} |u_{\vec k_{*}\alpha}|^{2}(\partial_{i} u_{\vec k_{*}\alpha} \partial_{j} \overline{u}_{\vec k_{*}\alpha} + \partial_{j} u_{\vec k_{*}\alpha} \partial_{i} \overline{u}_{\vec k_{*}\alpha}  \nonumber 
    \\
    + & 
    u_{\vec k_{*}\alpha} \partial_{i} \partial_{j} \overline{u}_{\vec k_{*}\alpha}  +
 \overline{u}_{\vec k_{*}\alpha} \partial_{i} \partial_{j} u_{\vec k_{*}\alpha}) +    \frac{1}{2} \partial_{i} \partial_{j} \varepsilon_{0}(\vec k_{*}) . 
 \nonumber
\end{align}
Here we included the dispersion-related terms  $\partial_{i} \varepsilon_{0}(\vec k_{*}) $ and $\frac{1}{2} \partial_{i} \partial_{j} \varepsilon_{0}(\vec k_{*})$  in the end of the definitions, since in the limit of the perfectly flat band this contribution vanishes and only the interaction-induced terms $\sim U n_0$ remain.

Finally, the tensor $\mathcal Y_{ij}$ in the expansion~\eqref{expansion} depends only on the Bloch functions, but does not depend on the quasiparticle dispersion,
 \begin{align}
   \mathcal Y_{ij} = &  2 U n_{0} \sum_{\alpha} \overline{u}_{\vec k_{*}\alpha}^{2} (u_{\vec k_{*}\alpha} \partial_{i}\partial_{j} u_{\vec k_{*}\alpha} -\partial_{i} u_{\vec k_{*}\alpha} \partial_{j} u_{\vec k_{*}\alpha}).    \nonumber 
\end{align}

By diagonalizing $\sigma_{z}\mathbb{H}_{\text{P}} (\vec q)$ in Eq.~\eqref{expansion}, we obtain the explicit dispersion relation for the Bogoliubov quasiparticles,
\begin{equation}\label{Bogoliubov}
    \omega (\vec q) = \mathcal V_{i} q^{i} + \sqrt{2 U n_{0} \Theta \, \chi_{ij}  \, q^{i}q^{j}} ,
\end{equation}
where the tensor $\chi_{ij}$ is given by
\begin{align}
    \chi_{ij} =  2 \mathcal W_{ij} - \mathcal Y_{ij} - \overline{\mathcal Y}_{ij} . 
\end{align}
and $\Theta \sim 1$ is determined by Eq.~\eqref{Theta}.
The tensor $\chi_{ij}$ is connected with both the quantum-geometric terms (represented by derivatives of the Bloch functions) and dispersive terms (represented by derivatives of the dispersion relation). 
We proceed to separate the contribution of the inverse effective mass tensor, denoted as $m^{-1}_{ij}$, from the tensor $\chi_{ij}$. This separation is expressed as
\begin{align}
 2 U n_{0} \tilde \chi_{ij} = \chi_{ij} - \partial_{i} \partial_{j} \varepsilon_{0}(\vec k_{*}),
\end{align}
where we introduce the factor $2 U n_{0}$ to ensure that the new tensor $\tilde \chi_{ij}$ in our units is dimensionless.
Hence, it is defined by 
\begin{align}
 \tilde \chi_{ij}  =    \sum_{\alpha} &  (|u_{\vec k_{*}\alpha}|^{2} u_{\vec k_{*}\alpha} \partial_{i} \partial_{j} \overline{u}_{\vec k_{*}\alpha}  + |u_{\vec k_{*}\alpha}|^{2} \overline{u}_{\vec k_{*}\alpha} \partial_{i} \partial_{j} u_{\vec k_{*}\alpha}
 \nonumber
 \\
 + & 2|u_{\vec k_{*}\alpha}|^{2} \partial_{i} u_{\vec k_{*}\alpha} \partial_{j} \overline{u}_{\vec k_{*}\alpha}  
  + 2|u_{\vec k_{*}\alpha}|^{2} \partial_{j} u_{\vec k_{*}\alpha} \partial_{i} \overline{u}_{\vec k_{*}\alpha}  \nonumber
 \\
 + &   \overline{u}_{\vec k_{*}\alpha}^{2} \partial_{i} u_{\vec k_{*}\alpha} \partial_{j} u_{\vec k_{*}\alpha} 
 + u_{\vec k_{*}\alpha}^{2} \partial_{i} \overline{u}_{\vec k_{*}\alpha} \partial_{j} \overline{u}_{\vec k_{*}\alpha}).
\end{align}
The tensor $\tilde \chi_{ij}$ plays a crucial role in determining the dispersion of Bogoliubov modes \eqref{Bogoliubov} and the speed of sound. Therefore, it warrants further examination. We note that this tensor is orbital-independent.

\textbf{Role of the quantum-geometric tensor}.  We can demonstrate the importance of the sound tensor $\tilde{\chi}_{ij}$ in capturing information about the quantum geometry of the underlying Bloch states. Let us focus on the BEC-i phase for clarity. In this phase, $\mathcal{V}_i$ simplifies to $\mathcal{V}_i = \partial_i \varepsilon_0(\vec{k}_*)$, which vanishes, since condensation occurs at the minimum of the dispersion. Additionally, for the BEC-i phase, we can use the identity $|\xi_{\alpha}|^{2} = 1/\mathcal{A}$ and obtain
\begin{align}
\tilde \chi_{ij}   = 
 \frac{1}{\mathcal A}  &   \sum_{\alpha} 
 u_{\vec k_{*}\alpha} \partial_{i} \partial_{j} \overline{u}_{\vec k_{*}\alpha}  
  + 2 \partial_{j} u_{\vec k_{*}\alpha} \partial_{i} \overline{u}_{ \vec k_{*}\alpha}   +  \text{c.c.} \nonumber 
 \\
 +  &  \sum_{\alpha}   \overline{u}_{ \vec k_{*}\alpha}^{2} \partial_{i} u_{ \vec k_{*}\alpha} \partial_{j} u_{ \vec k_{*}\alpha} 
 +  \text{c.c.}  
\end{align}
To eliminate terms with the second derivatives, we can use the relation
$
    \partial_{i} \partial_{j} \sum_{\alpha} |u_{\vec k_{*}\alpha}|^{2}  
    = 0 
$, resulting in
\begin{align}
  \tilde \chi_{ij}   = 
 \frac{2}{\mathcal A} \text{Re}  \sum_{\alpha} \left[ \partial_{i} u_{\vec k_{*}\alpha} \partial_{j} \overline{u}_{\vec k_{*}\alpha} + \mathcal{A}  \overline{u}_{\vec k_{*}\alpha}^{2} \partial_{i} u_{\vec k_{*}\alpha} \partial_{j} u_{\vec k_{*}\alpha} \right].  
 \nonumber 
\end{align}  
Next, we add and subtract  term  $\frac{2}{\mathcal A} 
    \Re \sum_{\alpha,\beta}$$\overline{u}_{\vec k_{*}\alpha} \partial_{i} u_{\vec k_{*}\alpha}$ $u_{\vec k_{*}\beta}\partial_{j} \overline{u}_{\vec k_{*}\beta}$, and arrive at
\begin{align}
    \tilde \chi _{ij}  = 
    \frac{2}{\mathcal A} \sum_{\alpha}  \mathfrak G_{ij} +  \frac{2}{\mathcal A} 
    \Re \left[ \sum_{\alpha,\beta}   \overline{u}_{\vec k_{*}\alpha} \partial_{i} u_{\vec k_{*}\alpha} u_{\vec k_{*}\beta}\partial_{j} \overline{u}_{\vec k_{*}\beta}
    \right. 
    \nonumber
    \\
    \left.
    +  \mathcal A \sum_{a} \overline{u}_{\vec k_{*}\alpha}^{2} \partial_{i} u_{\vec k_{*}\alpha} \partial_{j} u_{\vec k_{*}\alpha} \right] .
    \label{sound-tensor}
\end{align}
The quantity $\mathfrak{G}_{ij}$ in the first term of Eq.~\eqref{sound-tensor} represents the quantum-geometric tensor \cite{Provost1980}, which is defined in Eq.~\eqref{metrics}. This tensor provides the leading contribution to the speed of sound when dispersive effects can be neglected.

It is important to note that in addition to the quantum-geometric tensor $\mathfrak{G}_{ij}$, the sound tensor $\tilde \chi_{ij}$ may contain additional terms, which depend on the Bloch functions and their derivatives. Whether these terms vanish or not depends on the microscopic details and the symmetries of the effective Hamiltonian. For example, in the case of kagome lattice, these terms vanish, and the resulting Bogoliubov dispersion and speed of sound depend solely on the quantum metric $\mathfrak{G}_{ij}$ \cite{Julku2021}. This property arises from unique characteristics of the effective tight-binding model on the kagome lattice, where \textit{all hopping terms are real}, and each lattice site is the center of inversion.
These symmetries of the effective Hamiltonian lead to the Bloch states of the form $u_{\vec{k}*\alpha} \propto \overline{u}_{\vec{k}_*\alpha}$, allowing us to choose a gauge such that all the Bloch wave functions are real. Under this condition, the third term in Eq.~\eqref{sound-tensor} reduces to the quantum-geometric tensor, while the second term vanishes due to the equality
\begin{align}
    \sum_{\alpha} \overline{u}_{\vec k_{*}\alpha} \partial_{i} u_{\vec k_{*}\alpha} = \sum_{\alpha} u_{\vec k_{*}\alpha} \partial_{i} u_{\vec k_{*}\alpha} = \frac{1}{2} \partial_{i} \sum_{\alpha} u_{\vec k_{*}\alpha}^{2} = 0. \nonumber
\end{align} 
In this case, one obtains  
\begin{equation}
    \tilde \chi _{ij}  =  \frac{4 }{\mathcal A }  \mathfrak G_{ij} . 
\end{equation}
For the gapped topological system, represented here by the Haldane model \eqref{haldane1}, the presence of complex hopping terms introduces additional terms in the sound tensor~$\tilde \chi_{ij}$ beyond the quantum-geometric contribution. These terms can have impact on the Bogoliubov dispersion and the speed of sound. 

\begin{figure}[t]
 \includegraphics[width=\columnwidth]{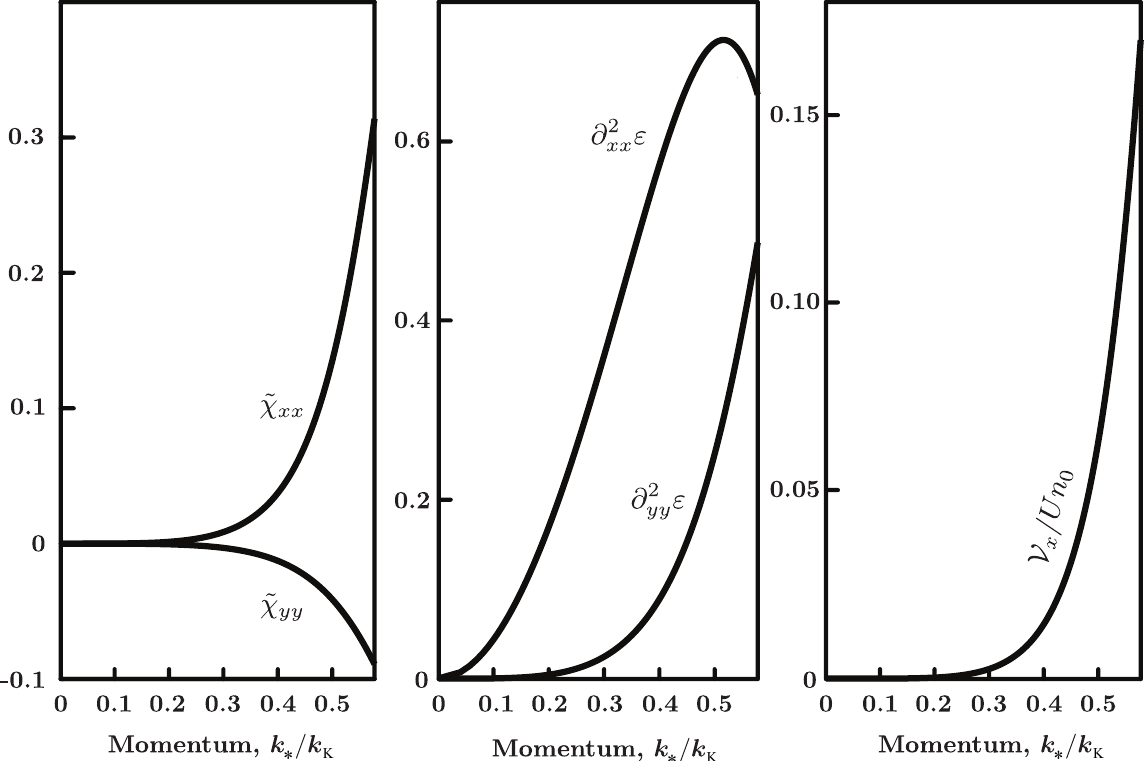} 
 \caption{\label{fig:Bogoliubov_Dispersions2}%
  \label{Fig-tensors} Dependencies of the main tensors determining the Bogoliubov dispersion and the speed of sound on the condensate momentum $k_*$ at $Un_0=0.2$ and $\Phi=\pi/5$.
  }
\end{figure}

Beyond any simplifying assumptions, the tensor $\tilde{\chi}_{ij}$ can be addressed only numerically. In Fig.~\ref{Fig-tensors} we visualize dependencies of the tensors  $\tilde{\chi}_{ij}$ and $\partial_{i} \partial_{j} \varepsilon_{0}(\vec k_{})$ defining the Bogoliubov dispersion on the condensate momentum~$\vec k_{*}$. 
The variation of $\vec k_{*}$ is realized here by the change of the Haldane hopping~$t_{2}$, assuming an implicit connection between $t_{2}$, $ U n_{0}$, and $\vec k_{*}$ through the mean-field minimization.
At small $|\vec k_{*}|$, it is evident that the tensor $\tilde{\chi}_{ij}$ is significantly smaller than the dispersive term. This is attributed to the proximity of the $\Gamma$-point, where only trivial geometric contributions exist. However, at larger wave vectors $|\vec k_{*}|$ and moderate $U n_{0}$, the geometric contribution can become comparable to the dispersive term.
It is important to note that the term $\mathcal{V}_{i}$ is generally rather small. Additionally, this term exhibits a relatively large nonlinear dependence on $U n_{0}$, which cannot be inferred from the lowest-band projected Hamiltonian. These nonlinear corrections typically result in $\mathcal{V}_{i}$ being even smaller than the values predicted by the lowest-band projected Hamiltonian.

\textbf{Influence of the higher Bloch bands}. The impact of the higher bands on the condensate wave functions and Bogoliubov modes requires a special attention. In the preceding discussion, we made an assumption that $\xi_{\alpha} = u_{ \vec k_{}\alpha} = u_{0 \vec k_{}\alpha}$. While this assumption holds for the phases BEC-i and BEC-ii, regardless of the value of $U n_{0}$, it remains valid only up to the leading order in $U n_{0} /\Delta$ for the BEC-iii phase. 
However, we cannot overlook the significance of these terms for the BEC-iii phase, since it exhibits unique characteristics that necessitate corrections to the condensate wave function.
To obtain these corrections from the higher bands, we introduce a modified expansion for the wave function $\xi_{\alpha}$ in the Gross-Pitaevskii equation,
\begin{align}
  \xi_{\alpha } = u_{0 \vec k_{*} \alpha} + \sum_{n>0} f_{n} u_{n \vec k_{*}\alpha} ,
\end{align}
where $n=0$ denotes the lowest band and $n>0$ labels the higher bands. 
We assume that the coefficients $f_{n}$ are of the order of $U n_{0}/\Delta$ and rearrange the terms accordingly. This leads to the modified wave function
\begin{equation}\label{HBdressing}
        |\xi \rangle= |u_{0 \vec k_{*}}\rangle -  \sum_{n>0} |u_{n \vec k_{*}}\rangle 
        \frac{2 U n_{0} \, \varkappa_{n0} (\vec k_*) }{\varepsilon_{n}(\vec k_{*}) - \varepsilon_{0}(\vec k_{*})}
        \end{equation}
with
\begin{equation}
 \varkappa_{n0} (\vec k_*) = \sum_{\alpha} \langle u_{n \vec k_{*} \alpha}| \ |u_{0 \vec k_{*}\alpha }|^{2}  \,  |u_{0 \vec k_{*} \alpha} \rangle .
\end{equation}
Here the leading-order renormalization vanishes in the BEC-i and BEC-ii phases, yet stays finite in the BEC-iii phase. 
By substituting the renormalization~\eqref{HBdressing} into the condition \eqref{minimum}, we obtain the renormalized relation for the quasiparticle velocity:
\begin{align}
 \partial_{i} \varepsilon_{0}(\vec k_{*}) + 2 U n_{0}  \left[\langle \partial_{i} u_{0 \vec k_{*}\alpha}| \mathcal P_{\alpha\beta}   
     |u_{0 \vec k_{*}\beta}\rangle +  {\rm c.c.} \right]  = 0,
     \label{velocity1}
\end{align}
 where  we introduced the operator
\begin{align} 
  \mathcal P_{\alpha\beta} = \left(1 - |u_{0\vec k_{*}}\rangle \langle u_{0\vec k_{*}}|\right)|u_{0,\vec k_{*}\alpha}|^{2} \delta_{\alpha \beta}.
\end{align}

It is important to note the distinction between the two terms in Eq.~\eqref{velocity1}. 
The first term represents quasiparticle velocity arising from the dispersive term, while the second term in Eq.~\eqref{velocity1} is reminiscent of \textit{anomalous velocity}, analogous to the one in the electronic systems \cite{Kruchkov2023,Haldane2004}. This term takes its origin from the nontrivial Berry connections of the underlying Bloch states, hence it is important for the BEC-iii phase, which arises from the interplay of small band width, nontrivial topology, and interactions. 
As a consequence of Eq.~\eqref{velocity1}, the quasiparticle velocity in the BEC-iii phase 
is \textit{augmented} by the topological effects scaled by the strength of interactions $U n_{0}$. This observation has been verified through numerical analysis.

Let us provide additional insights into the impact of higher-band corrections on the Bogoliubov excitations and the projected Hamiltonian. In the limit $\vec q \to 0$, one can project it onto the wave function $|\xi\rangle$. However, for the nonzero $\vec q$, we can achieve higher accuracy by projecting the Bogoliubov Hamiltonian onto the wave functions $|\xi_{\vec k_{} + \vec q}\rangle$. These wave functions are defined as the minimizers of the mean-field energy~\eqref{Energy} for $\vec k = \vec k_{} + \vec q$, or alternatively, according to Eq.~\eqref{HBdressing} with $\vec k_{}$ replaced by $\vec k_{} + \vec q$. The advantage of this improved projection method is clearly demonstrated in Fig.~\ref{fig:Bogoliubov_Dispersions} (labeled as P2). By employing the introduced projection procedure, we obtain the renormalized expressions for $\chi_{ij}$ and $\mathcal V_{i}$ which enter the Bogoliubov frequencies~\eqref{Bogoliubov}, yet these corrections are of the second order $\sim (U n_{0})^2/\Delta$.

\subsection{Speed of sound and effects of quantum geometry}

\textbf{Sound in BEC-i}. To gain insights from conventional condensates in cold gases with Eq.~\eqref{SoundSpeed} determining the sound velocity, let us revisit the trivial phase BEC-i. In this phase, condensation occurs at $\vec k_{*}=0$. The Bogoliubov dispersion \eqref{Bogoliubov} is determined by two quantities: ${\cal V}_{i}$ and $\chi_{ij}$. Due to the spatial homogeneity of the condensate, the quantity $\mathcal V_{i}$ vanishes. This can be understood by expressing the remaining term ($\sim U n_0$) of $\mathcal V_{i}$ as follows:
\begin{align}
\mathcal V_{i} = 4 U n_0 \sum_{\alpha} |u_{\vec k_{}\alpha}|^{2}(u_{\vec k_{}\alpha} \partial_{i} \overline{u}_{\vec k{}\alpha} + \overline{u}_{\vec k{}\alpha} \partial_{i} u_{\vec k_{}\alpha})  \nonumber 
\\
= \frac{4 U n_0}{\mathcal A} \partial_{i} \sum_{\alpha} |u_{\vec k_{*}\alpha}|^{2} = 0.
\nonumber 
\end{align}
Furthermore, in the case of condensation at $\vec k = 0$, it can be shown that all the first and second derivatives of the Bloch state $| u_{0,\alpha} \rangle$ vanish in a similar manner. Indeed, by expanding the Bloch Hamiltonian $\Ham^{\alpha\beta}(\vec k)$ around $\vec k = 0$, we obtain 
\begin{align}
\Ham^{\alpha\beta}(\vec k)  \simeq 3 t_{1} \left(1 - \frac{|\vec k|^{2}}{4} \right)\sigma_{x} + 2 t_{2} \cos{\Phi} \left(1 + \frac{9 |\vec k|^{2}}{4} \right) \sigma_{0},
\nonumber
\end{align}
where the Pauli matrices $\sigma_{0}$ and $\sigma_{x}$ act in the sublattice space. It is evident that the eigenvectors of $\Ham^{\alpha\beta}(\vec k)$ are independent of $\vec k$ up to $\mathcal O(\vec k^{3})$. Consequently, for the BEC-i phase, we have $\partial_{i} u_{\vec k_{} = 0\alpha} = \partial_{i} \partial_{j} u_{\vec k_{} = 0,\alpha} = 0$.

Our analysis yields a decisive conclusion: in the BEC-i phase, all quantum-geometric effects are completely nullified. Notably, the only residual contribution to the speed of sound emerges solely from the (inverse) effective mass. This contribution can be precisely described by the diagonal term 
\begin{align}
 m^{-1}_{ij}(\vec k_{*}) = (3 t_{1}/2 - 9 t_{2} \cos{\Phi}) \delta_{ij}.
\end{align}
These findings affirm the dominance of the inverse effective mass in shaping the behavior of the BEC-i phase, underscoring the simplicity of its characteristics.
Consequently, the Bogoliubov modes in BEC-i represent ordinary excitations of an isotropic and homogeneous BEC, characterized by 
\begin{align}
 \omega_{\vec q}  =  \sqrt{\frac{2U n_0 \Theta}{m} } q. 
\end{align}
Hence, the speed of sound in BEC-i phase is 
\begin{align}
 c  =  \sqrt{\frac{2U n_0 \Theta}{m} } , 
\end{align}
in agreement with the known result \eqref{SoundSpeed}.

\textbf{Sound in BEC-ii}.  The speed of sound in the BEC with high-symmetry momentum modulations has been extensively studied in Refs.~\cite{Julku2021, Julku2022}. Rather than providing a detailed discussion, we refer the reader to those references and offer a brief overview of the potential speed of sound characteristics and associated challenges. In Ref.~\cite{Julku2021} it was discovered that for certain gapless Hamiltonians, such as those on the kagome lattice, the speed of sound scales as $U n_0$ instead of $\sqrt{U n_0}$ as in ultracold gases \eqref{SoundSpeed}, given by
\begin{align}
 c \sim U n_0 \sqrt{ \text{Re} [\mathfrak G_{ii} (\vec k_*)] } . 
\label{BEC-ii-sound}
\end{align}
Additionally, the speed of sound probes the Fubini-Study metric (the real part of the quantum geometric tensor) at the wavelength of $2 \pi /k_*$. While this result is intriguing, it is not universal. Since the speed of sound is a macroscopic observable, it should be independent of orbital choice, yet the quantum metric at a particular momentum relies on the selection of orbital positions \cite{Julku2022}. Consequently, this relationship holds for the kagome lattice with real hopping but is not a general rule. In general, one should not expect a simple connection between quantum metrics and the speed of sound.

Moreover, as discussed in Sec.~\ref{sec:2}, there are additional challenges. The pure BEC-ii phase with the condensate wave vector $\vec k_{*}$ at the points K (K$'$) of BZ is unstable towards a phase where the condensate has wave vectors on different sublattices, as observed in the BEC-ii phase for the gapped Chern bands in the Haldane model. This instability manifests as instabilities at $\vec q = \vec K' - \vec K$, where three eigenvalues become negative instead of two, as indicated by the Bogoliubov quasiparticles. However, we do not delve into these details here, since an extensive ongoing numerical analysis is being conducted \cite{Salerno2023}.

\textbf{Sound in BEC-iii}.  In the BEC-iii phase, the condensate wave vector $\vec k_{*}$ generally differs from the minimum of the noninteracting dispersion relation due to the interplay between the interaction energy ${\cal U}$ and the dispersive contribution ${\cal K}$. Consequently, the first derivative of the dispersion relation, $\partial_{i} \varepsilon_{0} (\vec k_{*})$, is non-vanishing that results in a generally nonzero $\mathcal V_{i}$. This leads to a complex anisotropic behavior in the dispersion relation of the Bogoliubov excitations $\omega (\vec q)$ with respect to the wave vector~$\vec q$.
Due to the complexity of the general wave vectors~$\vec k_{*}$, here we do not provide analytical expressions for the tensors $\chi_{ij}$ and $\mathcal V_{i}$.

 \begin{figure}
 \includegraphics[width= 0.8  \columnwidth]{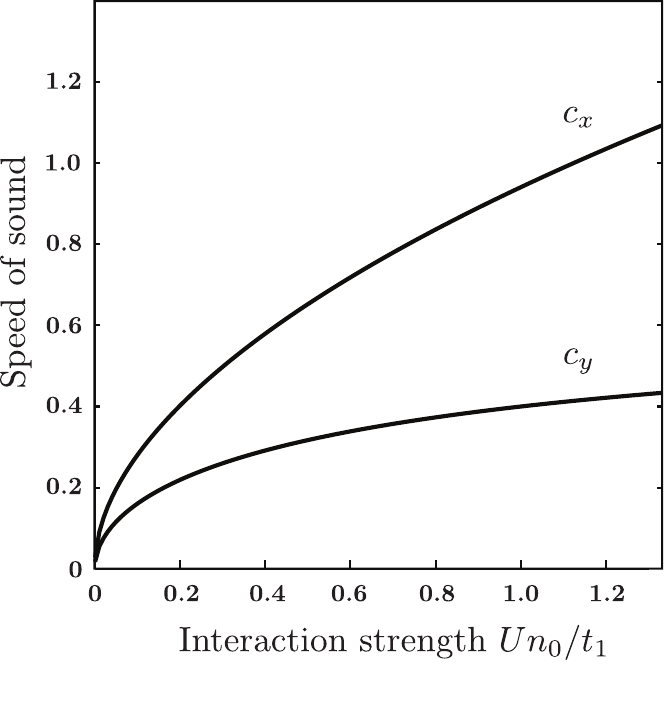} 
 \caption{\label{Fig-Sound}%
 Speed of sound (exact) as a function of the interaction strength in the BEC-iii phase at $\Phi=\pi/5$ and $t_{2} = 0.3$.
   }
\end{figure}
In Fig.~\ref{Fig-Sound} we show the numerical dependence of the sound velocity in the BEC-iii phase on the interaction strength $U n_{0}$.
The speed of sound, represented by the limit $\lim_{\vec q \to 0} \omega(\vec q)/|\vec q|$, exhibits complex anisotropic behavior. Specifically, we evaluate this quantity in the positive $x$ ($c_{x}$) and $y$ ($c_{y}$) directions.
We observe three distinct behaviors in the speed of sound for the BEC-iii phase:
(i) At small $U n_{0}$, the velocities $c_{x,y}$ scale as $\sqrt{U n_{0}}$ due to dispersive effects.
(ii) In the intermediate $U n_{0}$ regime, the velocity approximately follows the linear lependence on $U n_{0}$.
(iii) At larger $U n_{0}$, the higher-band effects introduce nonlinear corrections that lead to significant deviations between the exact Bogoliubov Hamiltonian and its lowest-band projection. 
In this regime, the lowest-band projection predicts a clear linear dependence of the velocity on $U n_{0}$.

\textbf{Remarks on the speed of sound.}  Through the numerical analysis, we observe notable enhancements in the speed of sound, driven by the nontrivial quantum metrics of the underlying bands. Although the derivation of analytical expressions remains challenging, these findings motivate us to investigate the implications of an augmented speed of sound on the superfluid's properties of the new superfluid BEC-iii. As an example, in the case of conventional superfluids, represented by the BEC-i phase, there exists a direct relationship between the speed of sound $c$ and the superfluid weight $D_S$, given by 
\begin{align}
 D_S \propto c^2 .
\end{align}
Consequently, the qualitative increase in the speed of sound, even within a narrow band scenario, amplifies the superfluid weight and subsequently elevates the Berezinskii-Kosterlitz-Thouless temperature. To delve deeper into this phenomenon specifically for BEC-iii, we conduct an in-depth exploration of the superfluid weight in topological bosonic bands, utilizing the powerful framework of the \textit{Popov hydrodynamic theory} for superfluids.

\section{Superfluid weight and quantum geometry}\label{sec:4}

When considering the superfluid state of bosons, an important quantity to examine is the superfluid weight, denoted as $D_S$. Before delving into detailed calculations later in this section, let us first discuss the physical significance and relevance of the superfluid weight. We begin with a conventional system where particles of mass $m$ form a superfluid with a density $n_0$. In this scenario, the superfluid weight is given by the expression:
\begin{align}
  D_0 (T) = \frac{n_0 (T) }{m}   , 
   \label{DS_conv}
\end{align}
meaning that the heavier are the particles, the lower is the superfluid weight. In the case of a perfectly flat band, the conventional superfluid weight \eqref{DS_conv} diminishes to zero ($D_S = 0$) as the particle mass tends towards infinity ($m \to \infty$). However, as we will explore in detail in this section, this is not true when the underlying flat band exhibits topological characteristics or, more precisely, when the band possesses nontrivial quantum geometry.

Generically, the superfluid weight is a tensor $D^{S}_{ij}$ ($i,j = x,y$), which defines the supercurrent response  $j^{S}_i$ to the external vector potential $\vec A$ (analogous to \textit{the Meissner effect} in superconductors),
\begin{align}
j^{S}_i =  - D^{S}_{ij} A_j. 
\label{vector potential}
\end{align}
Throughout this section, we will maintain the indices $i$ and $j$ in our derivations. Certain symmetries, such as the crystallographic $C_3$ symmetry in 2D, result in an isotropic superfluid weight  $D^{S}_{xx}=D^{S}_{yy}\equiv D_S$.  
In 2D systems, the superfluid weight $D_S$ holds great significance as it directly governs the BKT transition temperature,\footnote{For clarity, we will use $D^{S}_{ij}$ to denote the tensor of the superfluid weight, and $D_S$ to represent the scalar value in the specific isotropic case where $D^{S}_{xx}=D^{S}_{yy} \equiv D_S$.}
\begin{align}
 T_{\text{BKT} } \simeq \frac{\pi }{2}  \left[ \text{det} D^S_{ij} (0)  \right]^{1/2}, 
 \label{TBKT-est}
\end{align}
which serves as the critical point for the superfluid's breakdown in two dimensions.

From a field-theoretical standpoint, the emergence of the Bose-Einstein-condensed phase at low temperatures gives rise to the spontaneous breaking of the $U(1)$ symmetry. This, in turn, leads to the manifestation of gapless Goldstone bosons, which are manifested as the phase fluctuations $\phi$ of the field $c = \sqrt{\rho}e^{i\phi}$. These fluctuations play a pivotal role in low-energy physics and can be effectively described by the $XY$ model with the action \cite{Posazhennikova2006},
\begin{equation}
  S \equiv   \frac{D^{S}_{ij}}{2}  \, \frac{\partial  \phi } {\partial x_i} \, \frac{\partial  \phi } {\partial x_j} ,
 \label{SXY}
\end{equation}
where $D^{S}_{ij}$ is the tensor of superfluid weight.

In the framework of the linear response theory, the computation of the superfluid weight involves the evaluation of current-current correlators using the well-established quantum transport techniques (see the Kubo formalism in Ref.~\cite{Mahan2000}). Here, we employ the Matsubara technique in imaginary time $\tau$ to carry out the analysis. The superfluid weight, denoted as $D^{S}_{ij}$, can be determined through the Scalapino prescription \cite{Scalapino1993, Scalapino1992}:
\begin{equation}\label{Corr_func}
 D^{S}_{ij} = \langle \mathcal T_{ij} \rangle - \lim_{\vec q \to 0} \int_{0}^{\beta} d\tau \langle J_{i}(\vec q,\tau) J_{j}(-\vec q,0) \rangle_{c},
\end{equation}
where $\beta$ represents the inverse temperature, $\beta=1/T$. The operator $\mathcal T_{ij}$ is connected to the renormalized inverse effective mass tensor and is given by
\begin{align}\label{KuboT}
\mathcal T_{ij} = \sum_{\vec k} b^{\dagger}_{\vec k} \frac{ \partial^2 \Ham (\vec k) } {\partial {k_i} \partial {k_j} } b_{\vec k},
\end{align}
while the current operator $ J_{i}(\vec q)$ is expressed as
\begin{align}\label{KuboJ}
J_{i}(\vec q) = \sum_{\vec k} b^{\dagger}_{\vec k+\vec q} \frac{ \partial \Ham (\vec k+ \nicefrac{\vec q}{2}) }{ \partial k_i} b_{\vec k},
\end{align}
where $\Ham(\vec k)$ is the non-interacting Hamiltonian;  the impact of interactions can be studied by employing the Bogoliubov Hamiltonian upon the corresponding generalization. The correlation functions of these operators are then computed with the interacting Green's functions of the full model.

In this work, we employ the \textit{Popov hydrodynamic theory of superfluidity} as an alternative approach to derive an effective low-energy action for the phase fluctuations and compute the superfluid weight $D^{S}_{ij}$. The advantage of this approach is its applicability to nonperturbative Hamiltonians and its ability to circumvent infrared divergences.
Our analysis reveals that the superfluid weight of the bosonic superfluid in topological bands can be decomposed into two distinct contributions, denoted as 
 \begin{align}
  D^{S}_{ij} = D^{S0}_{ij}+ D^{\rm QG}_{ij}, 
 \end{align} 
 where $D^{S0}_{ij}$ is the conventional superfluid weight and $D^{\rm QG}_{ij}$ is the superfluid weight originating from nontrivial topology and quantum geometry (QG). 
In the remainder of the paper, we thoroughly investigate each contributing component of the superfluid weight and provide a comprehensive understanding of the underlying mechanisms.

\subsection{Bosonic superfluid weight $D^{S}_{ij}$ in the Popov theory of superfluidity}

The Popov hydrodynamic theory of superfluidity is a natural extension of the Bogoliubov model explored earlier \cite{Dupuis2009, Popov1979}. This approach adopts the phase-amplitude representation of bosonic fields, effectively circumventing infrared singularities. It specifically focuses on the low-momentum regime, known as the hydrodynamic regime, and offers valuable insights into the behavior of superfluids (see Ref.~\cite{Popov1987} and chapters 18--20 therein).

The hydrodynamic approach employed here involves a decomposition of the bosonic fields $b$ into the low-energy ($b^{\rm low}$) and high-energy ($b^{\rm high}$) modes. Notably, the high-energy modes exhibit behavior akin to nearly-free Bogoliubov quasiparticles within the effective field generated by the low-energy modes. The low-energy modes, on the other hand, possess a hydrodynamic nature, enabling their representation in terms of phase-density as $b^{\rm low} = \sqrt{n} e^{i\phi}$.
Subsequently, the integration of the high-energy modes leads to the definition of an effective action for the low-energy modes, specifically capturing the phase fluctuations. These fluctuations are effectively described by the $XY$ model, which plays a pivotal role in determining the superfluid weight as outlined in the definition~\eqref{SXY}.

\begin{figure}[t]
    \centering
    \includegraphics[width = 1.0 \columnwidth]{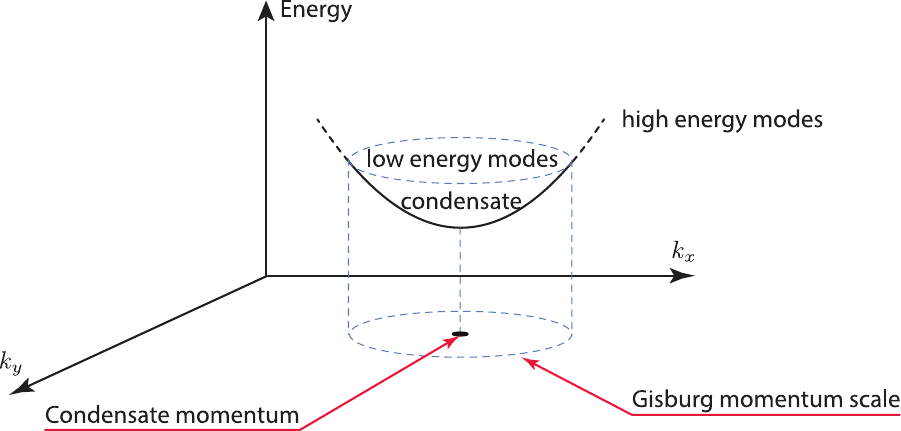}
    \caption{Schematic of separation into low energy modes and high energy modes in Popov theory of superfluids.}
    \label{Fig:Ginsburg}
\end{figure}

To define the low-energy modes for the flat band, it is important to consider the ``low-high energy'' decomposition with respect to the condensate. While one might initially assume that all modes of the flat band are low-energy modes, this is not the case. In our specific scenario, the Bogoliubov modes are gapped for the wave vectors $\vec{k}$ away from the condensate wave vector $\vec{k}_{*}$ (refer to Sec.~\ref{sec:2} for details).
Therefore, the low-energy modes are defined in the vicinity of the condensation momentum~$\vec{k}_*$, as illustrated in Fig.~\ref{Fig:Ginsburg}, satisfying
\begin{align}
|\vec q| = |\vec{k} - \vec{k}_{*}| < k_{\rm G},
\end{align}
where the momentum cut-off $\vec{k}_{\rm G}$ is referred to as the \textit{Ginsburg momentum scale} in the relevant literature \cite{Dupuis2009}. The value of $\vec{k}_{\rm G}$ is determined by assessing the smallness of the self-energy corrections $\Sigma(\vec{q} \sim \vec{k}_{\rm G}; \omega)$ in comparison to the Bogoliubov self-energies. This allows for the treatment of modes with $|\vec{q}| > k_{\rm G}$ as weakly interacting and effectively free.

We now turn our attention to the low-energy modes and adopt the phase-density representation. Specifically, for the site~$i$ with the coordinate $\vec{r}_i$ and sublattice index~$\alpha$, the low-energy mode can be expressed as%
\footnote{It is worth noting that we also consider a more general form $b^{\rm low}_{i \alpha} = \sqrt{n_{0}}\left(1+  \frac{\tilde \rho_{i}}{2}  \right) [\xi_{\alpha} + \delta \xi_{\alpha}(\phi, \rho)] e^{i \vec k_{*} \cdot \vec r_{i} + i \phi_{i}}$ as a possibility for better parametrization of the manifold of low-energy modes. Such deformations could introduce additional couplings between $\phi$ and $\rho$ and might be necessary to describe the presence of $\mathcal V$ in the Bogoliubov dispersion relation within the hydrodynamic framework. Furthermore, these deformations could result in small corrections (of the order of $\mathcal{V}^{2}/U$) to the superfluid weight for the new phase fluctuations. However, since the effects of these considerations are small for the phases under study, we do not delve further into this possibility.}
\begin{equation}
    b^{\rm low}_{i \alpha} = \sqrt{n_{0}}\left(1+  \frac{\tilde \rho_{i}}{2}  \right) \xi_{\alpha} e^{i \vec k_{*} \cdot \vec r_{i} + i \phi_{i}},
\end{equation}
where $n_{0}$ represents the condensate density and $\xi_{\alpha}$ corresponds to the sublattice component of the condensate wave function from the mean-field calculations. In this expression, $\tilde \rho_{i}$ denotes the density fluctuation, $e^{i \vec k_{*} \cdot \vec r_{i}}$ represents the spatial component of the condensate wave function, and $\phi_{i}$ represents a slowly varying phase fluctuation. To maintain consistency with the decomposition into the high-energy and low-energy modes, we limit the variation of $\tilde{\rho}_{i}$ and $\phi_{i}$ to small wave vectors in the vicinity of the condensate momentum, characterized by $|\vec k - \vec k_*| < k_{\rm G}$. 

We will now set aside the density fluctuations $\tilde \rho$ and revisit them later in this section.
For the \textit{high-energy modes} we use the representation\footnote{The phase factor $e^{i\phi_{i}}$ is essentially a convention, and it can be included in the redefinition of $\psi_{i\alpha}$.} 
\begin{equation}
   b^{\rm high}_{i\alpha} = \psi_{i \alpha} e^{i\phi_{i}}, 
\end{equation}
where $\psi_{i\alpha}$ corresponds to the high-energy field fluctuations with $|\vec k-\vec k_{*}| > k_{\rm G}$. This also includes the higher-band Bogoliubov fluctuations with arbitrary momenta. Hence,
\begin{equation}
    \sqrt{N}\phi_{i} = \phi_{\vec q} e^{i\vec q \vec r_{i}} + \phi_{-\vec q} e^{-i\vec q \vec r_{i}},
\end{equation}
where the real-valued phase $\phi_{i}$ implies that $\bar{\phi}_{\vec q} = \phi_{-\vec q}$ for the Fourier-transformed quantities.

\subsubsection{Effective action}

The superfluid action for the phase fluctuations $\phi_{\vec q}$ and high-energy modes $\psi_{\vec k \alpha}$ takes the following form with the defined low-energy and high-energy modes:
\begin{align}
S [  n_{0},  \psi_{\vec k\alpha }, \phi_{\vec q}] = S_{\rm MF}[n_{0}] + S_{\rm low}[n_{0}, \phi_{\vec q}]  & \nonumber 
\\
 +  S_{\rm high}[n_0, \psi_{\vec k \alpha}, \phi_{\vec q}] &,
\end{align}
The first term $S_{\rm MF}[n_{0}]$ in the effective action corresponds to the mean-field contribution, which determines the condensate wave vector and the densities $\xi_{\alpha}$ as discussed in Sec.~\ref{sec:2},
\begin{align}
    S_{\rm MF}[n_{0}] =  \beta N n_{0} \langle \xi|   \Ham( \vec k_{*}) |\xi \rangle - \beta N n_{0} \mu  &  \nonumber \\
      + \beta N U n_{0}^{2} \sum_{\alpha} |\xi_{\alpha}|^{4} & . 
\end{align}
The second term, denoted as $S_{\rm low}(n_{0}, \phi_{q})$, captures the interaction between the condensate and the low-energy phase fluctuations. It takes the form of a second-order expansion in terms of $\phi$ and $|\vec q|$, and can be expressed as follows:
\begin{align}
   S_{\rm low}[n_{0}, \phi_{\vec q}] = \beta n_{0} \phi_{\vec q} \phi_{-\vec q} q^{i} q^{j} \langle \xi| \partial_{i} \partial_{j} \Ham(\vec k_{*})|\xi\rangle.  
   \label{S-low}
\end{align}
The third term, denoted as $S_{\rm high}(\psi_{\vec k\alpha}, \phi_{\vec q}, n_{0})$, captures the interactions between the high-energy modes $\psi_{\vec k\alpha}$ and all the fluctuations, including both low-energy and high-energy components. In order to obtain the total superfluid weight $D^{S}_{ij}$ in the system, the high-energy modes $\psi$ need to be integrated out. This integration leads to the following expression for the corresponding action:
\begin{align}
    S_{\rm high }[n_0, \psi_{\vec k,a}, \phi_{\vec q}] = &   S^{j=0}_{\rm high 1} + \frac{1}{\sqrt{N}} S^{j=0}_{\rm high 2} +  \frac{1}{{N}} S^{j=0}_{\rm high 3 } 
    \nonumber 
    \\
    +  &  S^{j>0}_{\rm high }  + \mathcal O [\psi^3] . 
    \label{S-high}
 \end{align}   
In the given expression, the terms $S^{j=0}_{\rm high}$ represent the contribution from the lowest band, while the terms $S^{j>0}_{\rm high}$ describe the contribution from the upper bands of Bogoliubov quasiparticles with $|\vec k-\vec k_{*}|< k_{\rm G}$.  It is worth noting that the term $S^{j>0}_{\rm high }$ does not affect the contribution of other modes and will be discussed toward the end of our analysis. The terms of the order of $\psi^3$ and higher, denoted as $O(\psi^{3})$, account for the higher-loop corrections to self-energies and correlation functions, which are neglected in the present analysis.

The first term in the high-energy sector \eqref{S-high} characterizes the behavior of the $\psi$ modes in the noninteracting regime, resembling the Bogoliubov quasiparticles,
\begin{align}
  S^{j=0}_{\rm high 1} =    -  \sum_{i \omega_n} \sum_{|\vec k'-\vec k_{*}|>k_{\rm G}} \bar{\Psi}_{\vec k'} \, \left[i\omega_{n} \sigma_{z} - \mathbb{H}(\vec k') \right] \Psi_{\vec k'},
\end{align}
where $\Psi_{\vec k'} = [\psi_{\vec k' \alpha}; \psi^{\dagger}_{2 \vec k_{*} -\vec k'\alpha}]^{T}$ and the Bogoliubov Green's function is
\begin{align}
G^{-1} (i \omega_{n},\vec k) = i\omega_{n} \sigma_{z} - \mathbb{H} (\vec k). 
\end{align}
In this expression, $\omega_n = 2 \pi n/\beta$ represents the bosonic Matsubara frequencies. The summation over $\vec k$ is limited to a half of the Brillouin zone and indicated as $\vec k'$.

The second term can be expressed as follows:
\begin{align}
  S^{j=0}_{\rm high 2} =   i \mathbb{J}_{\vec q}  - i \mathbb{J}_{-\vec q} , 
\end{align}
where
\begin{equation}
        \mathbb{J}_{\vec q} = \int \limits_{0}^{\beta} d \tau  \, \phi_{\vec q} q_{i}  \sum_{|\vec k-\vec k_{*}|>k_{\rm G}} \psi^{\dagger}_{\vec k+\vec q}(\tau)\partial_{i} \Ham(\vec k+ \nicefrac{\vec q}{2}) \psi_{\vec k}(\tau) .
        \nonumber
\end{equation}
The third term can be written as 
\begin{equation}
      S^{j=0}_{\rm high 3}   = \int \limits_{0}^{\beta} d\tau \, \phi_{\vec q} \phi_{-\vec q} q_{i} q_{j} \sum_{ |\vec k-\vec k_{*}|>k_{\rm G}}  \psi^{\dagger}_{\vec k}(\tau) \partial_{i} \partial_{j} \Ham(\vec k) \psi_{\vec k}(\tau).   
      \nonumber
\end{equation}

We can achieve further simplification by explicitly summing the second and third terms,
\begin{align}
  S^{j=0}_{\rm high 2,3} =  \frac{1}{N} \int \limits_{0}^{\beta} &   d\tau        q_{i} q_{j}  \phi_{\vec q} \phi_{-\vec q} \mathcal T_{ij}(\tau)     
    \nonumber 
    \\ 
   +\frac{i}{\sqrt{N} }  \int \limits_{0}^{\beta} &   d\tau  \left.   q_{i} \left[ \phi_{\vec q}   J_{i}(\vec q,\tau)  - \phi_{-\vec q} J_{i}(- \vec q,\tau)  \right]  \right\},
\end{align}
The operators $\mathcal T_{ij}$ and $J_{i}(q)$ are defined in a similar manner to the linear response analysis, as shown in Eqs.~\eqref{KuboT} and \eqref{KuboJ}, 
\begin{align}
  \mathcal  T_{ij} = \sum_{|\vec k-\vec k_{*}|>\vec k_{\text G}}  \psi^{\dag}_{\vec k} \, \partial_{i} \partial_{j} \Ham(\vec k) \, \psi^{}_{\vec k},  
  \label{Tij}
\end{align} 
and
\begin{align}
   J_{i}(\vec q) = \sum_{|\vec k-\vec k_{*}|>\vec k_{\text G}} \psi^{\dag}_{\vec k+\vec q}\partial_{i}  \,
    \Ham(\vec k+ \nicefrac{\vec q}{2})  \,
    \psi^{}_{\vec k}.  
\end{align} 

We can integrate out all the higher modes $\psi_{\vec k}$, leading to the definition of an effective action $S_{\rm eff}(\phi_{\vec q})$ specifically for the phase fluctuations $\phi_{\vec q}$, 
\begin{align}
 S_{\rm eff} [\phi_{\vec q} ] \simeq    -\ln\Biggl\{\int \mathcal D \Psi_{\vec k} \  \exp \left[ {S^{j=0}_{\rm high 1}+ S^{j=0}_{\rm high 2,3}}   \right]  \Biggr\}.
 \label{S-eff}
\end{align}
Finally, a detailed analysis of contributions from the amplitude fluctuations shows that they can be considered as higher-order terms, which can be neglected in the current study.%
\footnote{More specifically, the action associated with the amplitude fluctuations $\tilde \rho$ takes the form:
\begin{align}
    S^{\tilde \rho}_{\rm high}  =  & i \beta n_0 q_i [\tilde \rho_{-\vec q} \phi_{\vec q} - \tilde \rho_{\vec q} \phi_{-\vec q}]  \langle \xi|\partial_{i} \Ham(\vec k_{*})|\xi \rangle 
 + 2 Un_{0}^{2} \beta \Theta \tilde \rho_{\vec q} \tilde \rho_{-\vec q}  . \nonumber
\end{align}
Furthermore, we utilize the property of the condensate to ensure that the condensate velocity at the momentum $\vec k_*$ is nullified,
\begin{align}
 \langle \xi| \partial_{i} \Ham(\vec k_{*})|\xi\rangle  \equiv 0.  
 \nonumber
\end{align}
Consequently, the coupling between the amplitude and phase fluctuations can be described as
\begin{align}
  S^{\tilde \rho}_{\rm high}  =  2 Un_{0}^{2} \beta \Theta \tilde \rho_{\vec q} \tilde \rho_{-\vec q}  \sim \mathcal O (\tilde \rho^2) . 
  \nonumber
\end{align}
In other words, the contribution from the amplitude fluctuations can be considered as higher-order terms, which can be neglected in the present analysis. However, it is important to note that these higher-order terms should be taken into account when studying phenomena such as sound damping effects \cite{Slyusarenko2012} or other problems involving hydrodynamic perturbation theory.}

With the derived effective action \eqref{S-eff} in hand, we proceed to calculating the superfluid weight in topological bosonic bands, with a specific focus on the new superfluid phase, BEC-iii.

\subsection{Conventional superfluid weight and its cancellation in the flat bands}

\textbf{Conventional superfluid weight}.  In this subsection, we derive the superfluid weight on the lattice \eqref{DS_conv} from the effective action.   Gaining a deeper understanding of the conventional term in the Popov approach involves distinguishing between the contributions from the lowest and higher energy bands,  
\begin{align}
 D^{S0}_{ij} = D^{S01}_{ij}+ D^{S02}_{ij}.  
\end{align} 

The first contribution, arising from the \textit{low-energy} modes of the lowest energy band \eqref{S-low}, can be obtained by evaluating the expression based on the definition \eqref{SXY},
\begin{align}\label{D1}
    D^{S01}_{ij} = n_{0}\langle \xi| \partial_{i} \partial_{j}
    \Ham(\vec k_{*})|\xi\rangle.
\end{align}
This expression bears resemblance to the conventional superfluid weight of the uniform condensates in ultracold gases \eqref{DS_conv}, since it incorporates the notion of the inverse effective mass. However, there appear additional contributions arising from the Bloch states on the lattice. We will delve into this connection in greater detail later on.

Let us analyze the influence of the low-energy Bogoliubov modes, originating from the \textit{higher bands}, on the superfluid weight. First, we define the higher-band Bogoliubov modes mathematically. If the eigenvalues of the operator $\tilde{\mathbb{H}}(\vec{k}) = \sigma_z \mathbb{H}(\vec{k})$ are given by $|n, s, \vec{k}\rangle$ (with the corresponding eigenvalues $sE_{n,s,\vec k_* + s(\vec k-\vec k_*)}$), where $n$ represents the band index and $s$ denotes the particle-hole index, we focus on the higher-band Bogoliubov modes with $n>0$ and express the field as
\begin{align}
\Psi_{\vec k_{}+ \vec q} = \sum_{s,n>0} \Psi_{n,s,\vec k_{}+\vec q} |n,s,\vec k_{}+\vec q\rangle .
\end{align}
The fluctuations in the form of $|0,s,k\rangle$ represent phase fluctuations and they are not included in $\Psi_{\vec k_{}+ \vec q}$.
We can then express the action $S^{n>0}_{\rm high}$ as
\begin{align}
 \frac{1}{\beta}  S^{n>0}_{\rm high 1 }  = 
 -  \sum_{i \omega_m }\sum_{s,n>0}  & \Psi^{\dagger}_{n,s,\vec k_{*}+\vec q,m} s(i\omega_{m} - sE_{n,s,\vec k_{*}+s\vec q}) 
 \nonumber 
 \\
 \times 
&  \Psi_{n,s,\vec k_{*}+\vec q,m} ,
   \nonumber
\end{align}
\begin{align}
 \frac{1}{\beta}  S^{n>0}_{\rm high 2 } =  i   \sqrt{n_{0}} q^{i} \sum_{s,n>0}  \bigl[ -  \phi_{-\vec q} \langle \bar{\Phi}_{0}|
   \hat   V^{i}_{\vec k_{*}+\vec q/2} 
   |n,s,\vec k_{*}+\vec q\rangle 
   \nonumber 
   \\
   \times \Psi_{n,s,\vec k_{*}+\vec q,0} 
   +  \phi_{\vec q}   \langle n,s,\vec k_{*}+\vec q|
   \hat   V^{i}_{\vec k_{*}+\vec q/2} 
   |\Phi_{0}\rangle \Psi^{\dagger}_{n,s,\vec k_{*}+\vec q,0} \Bigr],
   \nonumber
\end{align}
where we introduced the Bogoliubov velocities as
\begin{align} 
 \hat   V^{i}_{\vec k}  =  \sigma_{z} \partial_{i} \mathbb{H}(\vec k) .
 \label{velocity}
\end{align}

\begin{figure}[t]
  \includegraphics[width= 0.5 \columnwidth]{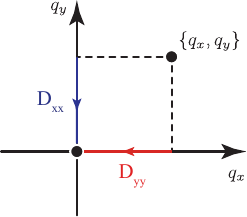} 
 \caption{Due to a response to the vector potential~\eqref{vector potential}, the calculation of limits $\vec q \to 0$ in superfluids  should be taken carefully. In particular, the Scalapino prescription  \cite{Scalapino1992,Scalapino1993} yields $q_{x}=0$, $q_y\to0$ for $D_{xx}$ and $q_{y}=0$, $q_x\to0$ for $D_{yy}$.  }
 \label{Scalapino}
 \end{figure}

 Upon integrating over the modes $\Psi_{n,s,\vec k_{*}+\vec q}$ with $n>0$, the expression for  the superfluid weight $D^{S02}_{ij}$ is obtained as follows:
\begin{align}
 \frac{  D^{S02}_{ij} }{n_0} = -  \sum_{s, n>0} 
    \frac{ \langle \Phi_{0}| \hat V^i_{\vec k_{*}+\vec q/2}|\Phi_{n,s}^{\vec q} \rangle  \langle \Phi_{n,s}^{\vec q} | \hat V^j_{\vec k_{*}+\vec q/2}|\Phi_{0} \rangle}{E_{n,s,\vec k_{*}+s\vec q}},
    \label{D2}
\end{align}
where $|\Phi_{n,s}^{\vec q} \rangle \equiv   |n,s,\vec k_{*}+\vec q \rangle $.  Equation \eqref{D2} can be rewritten in a more convenient form as follows:
\begin{multline}
  \frac{ D^{S02}_{ij}}{n_0} = \lim_{\vec q \to 0} \langle \Phi_{0}|\hat V^i_{\vec k_{*}+\vec q/2}   G_{\omega=0,\vec k_{*}+\vec q} \hat V^i_{\vec k_{*}+\vec q/2}  |\Phi_{0}\rangle,
  \label{DS2}
\end{multline}
where we evaluate the limit $\vec q \to 0$ according to the \textit{Scalapino prescription}  for the superfluid weight \cite{Scalapino1992,Scalapino1993}, see Fig.~\ref{Scalapino}. Below we demonstrate the equivalence between Eqs.~\eqref{DS2} and \eqref{D2}.
Indeed, the expression in question incorporates contributions from both the gapped and gapless Bogoliubov modes. Specifically, the contribution of the \textit{gapped modes} to the right-hand side of Eq.~\eqref{DS2} can be rewritten as the sum given in the right-hand side of Eq.~\eqref{D2} that results in a full equivalence of these two equations.

To address the contribution of the gapless modes, we introduce a projection onto the lower Bogoliubov band, using the renormalized lower band wave function $\xi_{\vec k_{*}+\vec q}$. After the projection, we can replace $G(0, \vec k_{*}+\vec q)$ with $-\mathbb{H}_{\text{P}} (\vec q)^{-1}$, and project $\hat V^i_{\vec k_{*}+\vec q/2}|\Phi_{0} \rangle$ onto the state $\xi_{\vec k_{*} + \vec q}$. This projection yields the vector
\begin{align}
 \begin{bmatrix}
  \langle \xi_{\vec{k_{*}}+\vec{q}}|
  \\
  \langle \overline{\xi}_{\vec{k}_{*}-\vec{q}}|
 \end{bmatrix}
 \hat V^i_{\vec k_{*}+\vec q/2}|\Phi_{0} \rangle = 
 \begin{bmatrix}
  \langle \xi_{\vec{k_{*}}+\vec{q}}|\partial_{i} \Ham_{\alpha \beta}(\vec k_{*} + \vec q/2) |\xi_{\vec k_{*}} \rangle
  \\
\langle \overline{\xi}_{\vec{k}_{*}-\vec{q}}| \partial_{i} \Ham^*_{\alpha \beta}(\vec k_{*} - \vec q/2) |\overline{\xi}_{\vec k_{*}} \rangle 
 \end{bmatrix}
 \nonumber
 \\
 = \begin{bmatrix}
  \langle \xi_{\vec{k_{*}}+\vec{q}}|\partial_{i} \Ham_{\alpha \beta}(\vec k_{*} + \vec q/2) |\xi_{\vec k_{*}} \rangle
  \\
  \langle \xi_{\vec{k}_{*}}| \partial_{i} \Ham_{\alpha \beta}(\vec k_{*} - \vec q/2) |\xi_{\vec k_{*}-\vec q} \rangle 
 \end{bmatrix}.
 \nonumber 
\end{align}
Expanding the expression in powers of the wave vector~$\vec q$, we find that to the leading order, the result is given by
\begin{align}
 \begin{bmatrix}
  \langle \xi_{\vec{k_{*}}+\vec{q}}|
  \\
  \langle \overline{\xi}_{\vec{k}_{*}-\vec{q}}|
 \end{bmatrix}
 \hat V^i_{\vec k_{*}+\vec q/2}|\Phi_{0} \rangle = 
 \langle \xi| \partial_{i} \Ham(\vec k_{*})|\xi \rangle  \begin{bmatrix}
  1
  \\
  1
 \end{bmatrix}
 + v_i
  \begin{bmatrix}
  1
  \\
  -1  
 \end{bmatrix} ,
 \label{Expansion}
 \nonumber 
\end{align}
where $v_i=m^{-1}_{ij} { q_j}/{2}$ and we make use of the definition of the effective mass tensor $m_{ij}$, 
\begin{equation}\label{mass}
   m^{-1}_{ij}  \equiv   
    \frac{ \partial^2 \varepsilon_{0}  (\vec k) } { \partial k_i \partial k_j}   \Bigr|_{\vec k = \vec k_*}.  
\end{equation}
The first term in this expansion vanishes due to the condition~\eqref{minimum}. Therefore,
\begin{align}
  \begin{bmatrix}
  \langle \xi_{\vec{k_{*}}+\vec{q}}|
  \\
  \langle \overline{\xi}_{\vec{k}_{*}-\vec{q}}|
 \end{bmatrix}
 \hat V^i_{\vec k_{*}+\vec q}|\Phi_{0} \rangle = 
 \frac{ 1}{2}   m^{-1}_{ij} q_j
  \begin{bmatrix} 
  1
  \\
  \text{-}1  
 \end{bmatrix} .
\end{align}
It is evident that the primary contribution to the expectation value in the limit $\vec q \to 0$ arises from the vectors $[1,-1]$.%
\footnote{Indeed, one obtains the matrix elements of $\mathbb{H}^{-1}_{\text{P}} (\vec q)$:  
 $$[1, -1] \mathbb{H}^{-1}_{\text{P}} (\vec q) [1, -1]^{T}  =  {8 U n_{0} \Theta}/ {\text{det} \mathbb{H}_{\text{P}} (\vec q)} \sim {1}/{q^2}, $$ while 
 $$
 [1, 1] \mathbb{H}^{-1}_{\text{P}} (\vec q) [1; - 1]^{T}
    = {-2 \mathcal V_{\sigma} q^{\sigma}}/{\text{det} \mathbb{H}_{\text{P}} (\vec q)} \sim {1}/{q},
 $$
 and 
 $$
[1, 1] \mathbb{H}^{-1}_{\text{P}} (\vec q) [1; 1]^{T}   = {\chi_{p\sigma} q^{p} q^{\sigma}}/{\text{det} \mathbb{H}_{\text{P}} (\vec q)} \sim \mathcal O (1). 
 $$}
  Hence, the contribution of the gapless modes to Eq. \eqref{DS2} can be expressed as follows:
\begin{align}
  2 U n_{0} \Theta   \lim_{\vec q \to 0}
  \frac{m^{-1}_{ii'} m^{-1}_{jj'}} {\text{det} \mathbb{H}_{\text{P}} (\vec q)}  q_{i'} q_{j'} . 
  \label{gapless}
\end{align}
In the final step of the calculation, the \textit{Scalapino prescription} \cite{Scalapino1992,Scalapino1993} is employed to evaluate the superfluid weight. Due to the symmetry of the system, both $m_{ij}$ and the matrix limit in Eq.~\eqref{gapless} are diagonal. To calculate $D^{S2}_{xx}$, we employ the transverse limit $q_{x}=0$, $q_y
\to 0$, and similarly for $D^{S2}_{yy}$, we apply the limit $q_{y}=0$, $q_x \to 0$. This yields
\begin{align}
 \lim_{ \substack{q_x\to 0 \\ q_y=0} }
  \frac{m^{-1}_{yi'} m^{-1}_{yj'}} {\text{det} \mathbb{H}_{\text{P}} (\vec q)}  q_{i'} q_{j'}  =    \lim_{ \substack{q_y\to 0 \\ q_x=0} }
  \frac{m^{-1}_{xi'} m^{-1}_{xj'}} {\text{det} \mathbb{H}_{\text{P}} (\vec q)}  q_{i'} q_{j'}  = 0.   \nonumber 
\end{align}
Therefore, the contribution of the gapless modes to Eq.~\eqref{DS2} is zero, confirming that Eq.~\eqref{DS2} is identical to Eq.~\eqref{D2}.

\begin{table*}[t]
\caption{Summary of the superfluid weight contributions in flat topological bands.}\label{tbl:3}
\centering
\begin{tabular}{@{}l l  l  l  l @{}}
\toprule
  &   Nature &  Main Origin  & Topological \\
 \midrule
$D^{S0}_{ij}$   & low-energy modes & effective mass and dispersion  &   no  \\
$D^{\rm QG}_{ij}$    &  quantum fluctuations &  quantum geometry  &  yes  \\
 \bottomrule
\end{tabular}
\end{table*}  

\textbf{Cancellation of the topological and quantum-geometric terms in the conventional superfluid weight}.  Let us analyze the contributions of the conventional terms $D^{S01}$ and $D^{S02}$ in more detail.
The first term $D^{S01}$ is the most basic one, for dispersive bosons,
\begin{equation}
   D^{S01}_{ij}  = n_{0} \langle \xi| \partial_{i} \partial_{j} \Ham(\vec k_{*})|\xi\rangle,
\end{equation}
and its primary contribution comes from the \textit{effective mass} \eqref{mass}. For instance, for a single-band system with the dispersion $ \varepsilon_0 (\vec k)$, characterized by the effective mass~$m$, the superfluid weight is determined by the condensate density $n_0$, 
\begin{align}
  D_0 \equiv  n_{0} \partial_{i} \partial_{j} \varepsilon_0 (\vec k_{*}) = \frac{n_{0}}{m}  .
\end{align}
This corresponds to the conventional superfluid weight introduced earlier in the discussion \eqref{DS_conv}.

However, in the case of a topological band structure, there are additional (quantum-geometric) contributions to consider. These contributions arise from two sources.
The first source arises in the limit of weak interactions, where $U n_{0} \to 0$. The second source is the renormalization of $|\xi\rangle$ due to the presence of higher bands, as shown in Eq.~\eqref{HBdressing}. The resulting renormalization is of the order of $\Delta \times (U n_{0}/\Delta) \sim U n_{0}$, where the gap $\Delta$  comes from the matrix elements of $\partial_{i} \partial_{j} \Ham(\vec k_{*})$. 
In the limit $U n_{0} \to 0$, the renormalization to $D^{S01}_{ij}$ from quantum geometry is given by:
\begin{align}
  D^{S01}_{ij} = 
   D_0 + n_{0}\sum_{n>0}  \Delta_{0n} [\mathfrak G^{0n}_{ij} (\vec k_*) + \mathfrak G^{0n}_{ji} (\vec k_*)],
\end{align}
where $\Delta_{0n}=[ \varepsilon_{n}(\vec k_{*}) - \varepsilon_{0}(\vec k_{*})]$, the sum runs over the higher bands with the Bloch index $n>0$ (in our notations, $n=0$ labels the lowest band), and $\mathfrak G^{nm}_{ij} (\vec k)$ is the multi-orbital quantum-geometric tensor \cite{Ma2010,Kruchkov2022,Iskin2022},
\begin{align}
\mathfrak  G^{nm}_{ij} (\vec k) =     \langle \partial_{i} u_{n \vec k } | u_{m \vec k} \rangle 
\langle  u_{m \vec k } | \partial_{j} u_{n \vec k} \rangle  .
\end{align}
For a system with two flat topological bands separated by the energy gap $\Delta$, the formula simplifies to
\begin{align}
D^{S01}_{ij} = D_0 + 2 n_{0} \Delta \, \text{Re}[\mathfrak{G}^{0}_{ij}(\vec{k})],
\end{align}
where $\mathfrak{G}^{0}_{ij}(\vec{k})$ is the quantum-geometric tensor \eqref{metrics} of the lowest band.

The second term  $D^{S02}_{ij}$ [see Eq.~\eqref{D2}] describes the back-reaction of the  Bogoliubov excitations of the higher bands to the phase fluctuations. This term is relevant for the \textit{dispersive bosons}, since its main contribution arises from the velocity operator~\eqref{velocity1},
 \begin{align}
  {  D^{S2}_{ij} } = - {n_0} \sum_{s, n>0} 
    \frac{ \langle \Phi_{0}| \hat V^i_{\vec k_{*}+\vec q/2}|\Phi_{n,s}^{\vec q} \rangle  \langle \Phi_{n,s}^{\vec q} | \hat V^j_{\vec k_{*}+\vec q/2}|\Phi_{0} \rangle}{E_{n,s,\vec k_{*}+s\vec q}}  , 
    \nonumber 
 \end{align}
In the limit of weak interaction $Un_{0} \to 0$, this term contributes as
\begin{align}
 {  D^{S02}_{ij} } = - {n_0} \sum_{n>0}  \Delta_{0n}  [\mathfrak G^{0n}_{ij} (\vec k_*) + \mathfrak G^{0n}_{ji} (\vec k_*)]. 
\nonumber
\end{align}
Hence, to the first order in the interaction strength, the quantum-geometric contributions to $D^{S1}_{ij}$ and $D^{S2}_{ij}$ cancel each other. For $U n_0 \ll \Delta$, their sum is given by
\begin{align}
  D^{S01}_{ij} + D^{S02}_{ij}  \simeq n_{0}\partial_{i}\partial_{j} \varepsilon_{0}(\vec k_{*}) \equiv D_0. 
\end{align}
It is worth noting that the conventional superfluid weight $D^{S0}_{ij} = D^{S01}_{ij} + D^{S02}_{ij}$ does not exhibit quantum geometrical or topological effects at the leading order. Therefore, in Table~\ref{tbl:3}, we refer to $D^{S0}_{ij}$ as nontopological.

To illustrate this concept, let us examine the superfluid weight in the BEC-i phase of the Haldane model. In this regime, the superfluid weight can be approximated as
\begin{align}
D^{S0}_{ij} \simeq n_0 (3 t_{1}/2 - 9 t_{2} \cos{\Phi}) \delta_{ij}. 
\end{align}
Subsequently, the \textit{upper bound} for the BKT transition temperature in this phase is given by
\begin{equation}
    T_{\rm BKT}  \simeq \frac{\pi}{2} n_0 (3 t_{1}/2 - 9 t_{2} \cos{\Phi}). 
\label{Tbkt-1}
\end{equation}

The first observation is that both the superfluid weight and the BKT temperature are maximized in the limit $\Phi=\pi/2$ (the so-called particle-hole symmetric case in the band structure of the Haldane model), which is characterized by
\begin{equation}
    T^{\rm max}_{\rm BKT}  \simeq \frac{3 \pi n_0  t_1}{4} . 
\label{Tbkt-2}
\end{equation}
Indeed, in the Haldane model, the parameter $n_0 t_1$ sets the scale for both the superfluid weight and the BKT transition temperature.

The second observation is that a deviation from the particle-hole symmetric point in the noninteracting band structure by means of the change of the flux $\Phi$ leads to a reduction in both the superfluid weight and the BKT transition temperature until they completely vanish. In particular, for a particle-hole broken scenario with flat bands ($\Phi \approx \pi/5$), Eq.~\eqref{Tbkt-1} reveals that conventional superfluidity ceases to exist only when $t_2/t_1$ is adjusted close to 0.2. This finding is consistent with the phase diagram depicted in Fig.~\ref{phasediagram}, thus affirming the robust superfluid nature of the BEC-i phase.

\subsection{Superfluid weight from the higher-energy modes and quantum geometry}

As it is shown above, in the case of infinite effective mass and weak interactions, the conventional superfluid weight vanishes,
 \begin{align}
 D^{S01}_{ij} + D^{S02}_{ij}  = D_0  \to 0. 
 \end{align}
However, it is important to acknowledge that there is an additional contribution that has been  been overlooked  thus far, originating from the \textit{higher-energy modes} of the Popov theory. To account for this contribution, we expand the effective action \eqref{S-eff} up to quadratic terms in $\phi_{\vec q}$ and $|\vec q|$,
\begin{align}
    S_{\rm eff}[\phi_{\vec q}] 
 &    \simeq 
    \frac{\phi_{\vec q} \phi_{-\vec q} q_{i} q_{j} }{N }  
    \left[
    \int \limits_{0}^{\beta} d \tau \langle \mathcal T_{ij}(\tau) \rangle  
    \right.\nonumber
    \\ 
   &  \left. -  \int_{0}^{\beta} \int_{0}^{\beta}  d\tau_{1} d\tau_{2} \, 
    \langle J_{i}(\vec q,\tau_{1}) J_{j}(-\vec q,\tau_{2}) \rangle_{c} \right] .
 \end{align}
Using the definition \eqref{SXY}, we can directly derive the expression for the superfluid weight as follows:
\begin{align}
  N D^{\rm QG}_{ij} (\vec q) 
 = \langle \mathcal T_{ij} \rangle 
 - \Pi_{ij}  (\vec q).
\end{align}
Here $\mathcal T_{ij}$ is the familiar diamagnetic contribution of the \textit{higher-energy modes} beyond the Ginsburg scale $k_{\rm G}$ [see Eq.~\eqref{Tij}]
and $\Pi_{ij} (\vec q)$ represents the paramagnetic current-current correlator,
\begin{align}
 \Pi_{ij} (\vec q)  = \int \limits_{0}^{\beta}  
 d\tau \langle J_{i}(\vec q,\tau) J_{j}(-\vec q, 0) \rangle_{c}.
\end{align}
The subsequent calculation follows a similar procedure to that outlined in previous literature (see, e.g., Refs.~\cite{Rossi2021, Julku2021}). We provide here a brief overview of the key steps and refer the reader to those references for computational details. First, we define the Matsubara transformation of the propagators~$G (\tau) =  -\langle  \Torder \psi^{\dag}_{\vec k} (\tau)  \psi^{\dag}_{\vec k} (0) \rangle$ to bosonic Matsubara frequencies  $\omega_n = 2 \pi n /\ \beta$, i.e., 
\begin{align}
G (\tau) = \frac{1}{\beta} \sum_{i \omega'_n} e^{-i \omega'_n \tau}  G(i \omega'_n)  .  
\end{align}
In evaluating the expectation value of the diamagnetic term, it is useful to consider a correlator of the form
$
	\langle \psi^{\dag}_{\vec k}  (\tau ) \, \partial_{i} \partial_{j} \Ham(\vec k) \, \psi^{}_{\vec k} (0)  \rangle
$
and take the limit $\tau \to 0$. This procedure results in
\begin{align}
\langle   \mathcal  T_{ij} (0) \rangle   
= 
 -\frac{1}{2\beta} \sum_{i \omega_n} \sum_{|\vec k-\vec k_{*}|>k_{\rm G}}  \Trace  [ G(\vec k, i \omega_n)   \,     \partial_{i} \partial_{j}\mathbb H  (\vec k)].
\end{align}  
The calculation of the connected part of the current-current correlation function 
is performed by implementing the Wick's theorem \cite{Mahan2000}, which yields
\begin{align}
  \Pi_{ij} (\vec q)
    = \frac{1}{2\beta}\sum_{i \omega_n} \sum_{|\vec k-\vec k_{*}|>k_{G}}  \Trace 
&    \hat V^i_{\vec k+\vec q/2}
   G_{i \omega_{n},\vec k} \nonumber
  \\
  \times
&\hat V^j_{\vec k+\vec q/2}
  G_{i\omega_{n},\vec k+\vec q}  , 
  \nonumber 
\end{align}
Upon a closer examination, we find that the obtained expressions resemble Eq.~\eqref{Corr_func} with two important differences.
First, there is an additional contribution to the superfluid weight from $S_{\rm low}[\phi_{\vec q}, n_{0}]$, which must be included in the calculation.
Second, the correlation function is computed using the action of the Bogoliubov excitations rather than the full bosonic action. As a result, the substitution $b \to \sqrt{Nn_{0}}\delta_{\vec k,\vec k_{*}} + \delta b_{\vec k}$ is not considered when evaluating the current-current correlation function. This is because the \textit{condensate is explicitly excluded} from $\psi_{\vec k}$ due to the low-energy cutoff $|\vec k-\vec k_{*}|>k_{\rm G}$.
Considering these factors, we obtain the final expression for the $D^{S3}_{ij}$ contribution to the superfluid weight in the form
\begin{align}
    &D^{\rm QG}_{ij} =   -\frac{1}{2N \beta }  \sum_{i \omega_n}   \sum_{|\vec k-\vec k_{*}|>k_{G}}   \Trace{ G_{i\omega_{n},\vec k}}  \partial_{i} \partial_{j}\mathbb{H}(\vec k)   \nonumber 
    \\
     &-   \frac{1}{2N\beta}  \sum_{i \omega_n}   \sum_{|\vec k-\vec k_{*}|>k_{G}}   \Trace 
    \hat V^i_{\vec k+\vec q/2} G_{i \omega_{n},\vec k}
    \hat V^j_{\vec k+\vec q/2} G_{i\omega_{n},\vec k+\vec q}   .
    \label{D3}
\end{align}

\begin{figure}[t]
  \includegraphics[width= 0.9 \columnwidth ]{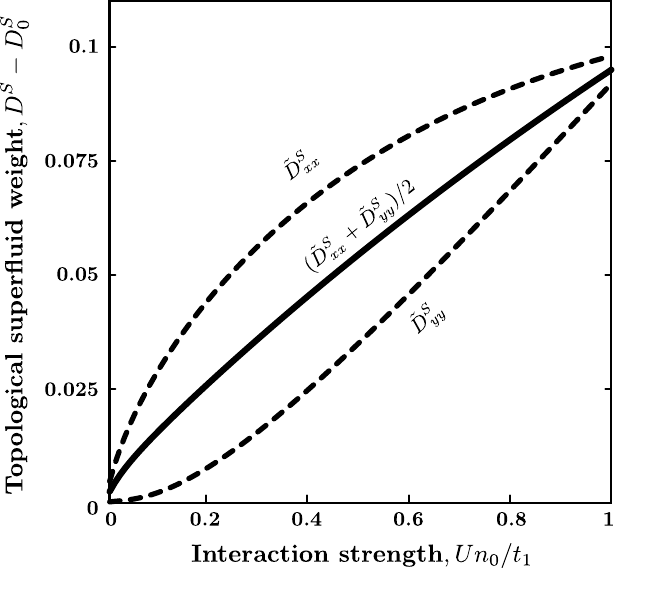} 
 \caption{
   Dependence of the superfluid weight $D^{S}$ with the subtraction of the trivial components $\partial_{i} \partial_{j}\varepsilon(\vec k_{*})$ on the interaction strength $U n_{0}$ at $n_{0} \approx 1$, $t_{2} = 0.28$, and $\Phi = \pi/5$ (BEC-iii phase). }
   \label{super-weight}
 \end{figure}

\textit{Numerical calculation of the superfluid weight}.  
We now proceed with the numerical calculation of the superfluid weight inside the BEC-iii phase, as summarized in Fig.~\ref{super-weight}. For nonzero $U n_{0}$ we obtain a number of the order of $U n_{0}$ corrections, which originate from the higher band corrections to the condensate wave functions. These corrections can lead to large changes in the sum $D^{S0}_{ij}$ at moderate $U n_{0} \approx 0.5$. 

First, we discuss the choice of the Ginsburg cutoff in Eq.~\eqref{D3}. It is well-known that the Bogoliubov approach encounters infrared (IR) divergences. To address these divergences, Popov introduced the cutoff parameter $k_{\rm G}$. In the Popov theory of superfluids, the IR divergences are effectively eliminated by treating the low-energy modes within the hydrodynamic perturbation theory. Therefore, the primary purpose of $k_{\rm G}$ in our calculation is to regulate these divergences.

By examining the leading terms in the perturbation theory, we observe that in our case they do not exhibit IR divergences. This suggests that, in principle, we can set $k_{\rm G}$ to zero in our calculation, as long as the region $|\vec k - \vec k_*| < k_{\rm G}$ contributes negligibly in the absence of IR divergences. In this scenario, the expression for the superfluid weight simplifies to the results obtained in the extended Bogoliubov approach, as developed in Ref.~\cite{Julku2021}. However, we acknowledge that there may be additional smaller corrections to the superfluid weight that are not considered here. Some higher-order corrections have been investigated recently in Ref.~\cite{Julku2022}, and it has been concluded that they do not significantly alter the numerical values.

To numerically calculate $D_{ij}^{\rm QG}$, we employ the spectral decomposition of the Green's function, which allows to take trace over eigenstates,
\begin{align}
 G (i\omega_{n}, \vec k_{*} + \vec q ) = \sum_{s} \sum_{i \omega_n}
 \frac{ s \, 
 |\Phi_{n \vec q}^{(s)}  \rangle  \langle \Phi_{n \vec q}^{(s)} | } { i\omega_{n} - sE_{n,\vec k_{*} + s\vec q} } .
\end{align}
Using these decompositions, we perform the Matsubara summation by employing contour integration \cite{Mahan2000}. The resulting expression then depends solely on the wave vectors $\vec k$, which are subsequently integrated over numerically, 
\begin{align}
    D^{\rm QG}_{ij} = \frac{1}{2} \sum_{\vec q} \sum_{nm} \sum_{ss'} s s'
    \frac{
    f( s E_{m,\vec k_* + s \vec q} ) - f( s' E_{n,\vec k_* + s' \vec q} )
    }
    {s' E_{n,\vec k_* + s' \vec q} - s E_{m,\vec k_* + s \vec q}
    }
    \nonumber
    \\
    \times
    \langle \Phi_{n \vec q}^{(s')}  | 
    \pi^- \partial_i \mathbb H (\vec q)    | 
    \Phi_{m \vec q}^{(s)}  \rangle 
    \langle \Phi_{m \vec q}^{(s)}  | 
    \pi^+ \partial_j \mathbb H (\vec q)
    |  \Phi_{n \vec q}^{(s')} \rangle .
    \label{DQG}
\end{align}
Here the integration is taken over $\vec k$ which is included in $\vec q = \vec k - \vec k_*$, $E (
\vec k)$ is the Bogoliubov dispersion, $f(E)$ is the Bose-Einstein distribution function, and $\pi^\pm=(\sigma_0 \pm \sigma_z)$. It is important to keep in mind that this expression contains an integrable divergence near the condensate wave vector $\vec k = \vec k_{*}$. Therefore, it is necessary to handle this divergence carefully during the integration procedure to obtain accurate results.

The presence of momentum derivatives in Eq.~\eqref{DQG} establishes a connection between this formula and the superfluid weight in fermionic systems \cite{Peotta2015, Rossi2021}. In particular, we observe that phases of nontrivial topology will have nonvanishing bosonic superfluid weight \eqref{DQG}, even if the dispersion is completely flat. However, in contrast to the fermionic case, where the superfluid weight is bounded from below by the Chern number, the bosonic problem described by Eq.~\eqref{DQG} exhibits additional levels of complexity. 
It does not seem  possible to reduce Eq.~\eqref{DQG} to a simple formula  bounding the  superfluid weight by the Chern number, yet such indirect relation might, in principle, exist. 
We plot the superfluid weight~\eqref{DQG} for a set of parameters, specifying the new superfluid phase BEC-iii, as given in Fig.~\ref{super-weight}. While the superfluid weight generally depends on the direction, the averaged superfluid weight in the sample, $D = (D_{xx}+D_{yy})/2$, shows an approximately linear scaling with the interaction strength given by
\begin{align}
D \propto U n_0, \nonumber
\end{align}
which is similar to the fermionic case as described in Ref.~\cite{Peotta2015}.

\subsection{Reentrant superfluidity}

We now discuss the qualitative behavior of the system by fixing the values of $t_1$, $t_2$ and varying the Haldane flux. This setting is motivated by  the framework of experiment~\cite{Jotzu2014}. To capture new effects related to the interplay between narrow bandwidth, strong interactions, and quantum geometry, we focus on the ratio $t_2/ t_1 \approx 0.3$, which minimizes the bandwidth, if the Haldane flux is tuned to $\Phi \approx \pi /5$. 

\begin{figure}[t]
  \includegraphics[width= 0.9 \columnwidth ]{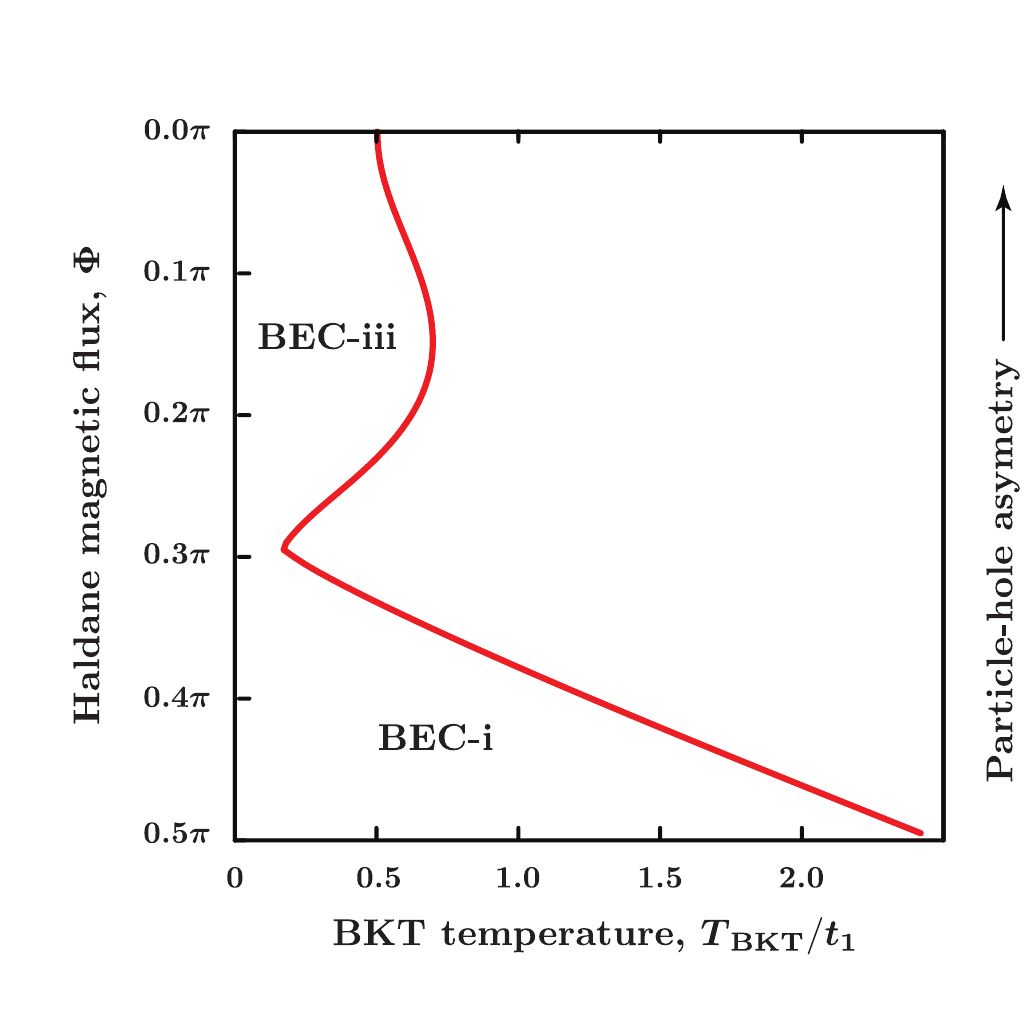} 
 \caption{Re-entrant superfluidity with the change of the Haldane magnetic flux: the upper bound of the BKT temperature (in units of $t_1$) at different values of the Haldane flux (here flux quantum = $2 \pi$). The temperature dependence is calculated from Eq.~\eqref{TBKT-est} for the Haldane model  \eqref{Ham0} with the parameters $t_2 = 0.28t_1$, $U = 0.5 t_1$, and $n_0 \approx 1$.  \label{Haldane-flux} }
 \end{figure}

The numerical results are depicted in Fig.~\ref{Haldane-flux} and summarize the dependence of the upper bound of the Berezinskii-Kosterlitz-Thouless temperature ($T_{\rm BKT}$) on the varying Haldane flux $\Phi$.
The flux ranges from $\Phi = \pi/2$ (particle-hole symmetric, dispersive, large band width) to $ \Phi =0$ (particle-hole asymmetric, flattened). We observe that as the band becomes flatter at intermediate values of  $\Phi$, the conventional contribution to the superfluid weight diminishes. Yet by further varying the Haldane flux, the condensation momentum~$k_*$ is pushed to the values where the quantum-geometric effects $\mathfrak G (\vec k_*)$ are relevant, driving the enhancement of the superfluid weight, and thus $T_{\rm BKT}$. Hence, we come to a conclusion that the re-entrant superfluidity in this case is promoted by the nontrivial interplay between quantum geometry, narrow bandwidth, and the interaction strength, with quantum geometry being an indispensible agent.

\textbf{ Experimental considerations:}  It is tempting to discuss a potential detection of the BEC-iii phase in setups such as ultracold gases in optical lattices modeling the Haldane model~\cite{Jotzu2014}.
The analysis of relevant energy scales indicates that the condition $t_2/t_1 \approx 0.3$ is reasonable. However, determining the accessible range of the Haldane phase becomes complicated, since it depends on both hoppings $t_2$ and $t_1$ ($[\Phi_{\rm min} (t_1, t_2), \Phi_{\rm max} (t_1, t_2)]$), as discussed in the Supplemental Information of Ref.~\cite{Jotzu2014}. 
A detailed analysis of the dependence $[\Phi_{\rm min} (t_1, t_2), \Phi_{\rm max} (t_1, t_2)]$  is beyond the scope of this study. Additionally, the challenge of heating in the experimental setup presents certain obstacles when analyzing experimental observables, especially when the effective temperature exceeds the bandwidth. This issue has not been addressed in the previous literature and requires further investigation. Nevertheless, the prospects of using the setup proposed in Ref.~\cite{Jotzu2014} for detecting the BEC-iii phase appear to be very promising.

Finally, we note that in condensed matter physics the re-entrant superconductivity is a rather exotic phenomenon, and it has recently been observed in the twisted trilayer graphene~\cite{Cao2021}, a two-dimensional material with flat bands in its dispersion~\cite{Carr2020, Khalaf2019}.
It is well established, that in its ``parent heterostructure'' twisted bilayer graphene, which  features topological flat bands~\cite{TKV} and nontrivial quantum geometry~\cite{Guan2022,Ledwith2020}, the superconducitvity is strongly enhanced by quantum-geometric contributions~\cite{Hazra2019,Hu2019,Xie2020, Julku2020, Torma2022};  for a recent experimental account of these effects see experimental study \cite{Tian2023}.
In this regard, it would interesting to determine whether a similar effect of quantum-geomtric contribution enhances the superfluid weight in the twisted trilayer graphene in a nonlinear way. Quantum geometry may or may not be responsible for  re-entrant supecronductivity in the scenario when the local phases are modified by external fields; we leave this question to future studies.

\section{Discussion}

In this study, we investigated the phenomenon of Bose-Einstein condensation and superfluidity in multiband lattice systems, highlighting the intricate interplay between interactions, dispersion relation, and quantum geometry. Specifically, we focused on examining the properties of superfluid phases in the context of the Bose-Haldane model with local on-site interactions.
Within our mean-field analysis, we have uncovered the emergence of several remarkable superfluid phases, including the exotic BEC-ii and BEC-iii phases. The BEC-ii phase exhibits condensation in multiple wave vectors, concentrated on different sublattices, while the BEC-iii phase is characterized by a variable condensate wave vector influenced by the dispersive properties of the lowest band, higher bands, the strength of interactions, and quantum geometry.

In the corresponding analysis of these phases, we studied the properties of the respective Bogoliubov excitations and developed a methodology to incorporate the higher-band corrections into the analytical calculations. Our results demonstrate that the interplay between dispersion relations, band geometry, and higher-band effects leads to the complex behavior in the velocities of Bogoliubov excitations, which depend on the interaction strength between particles in the way beyond Ref.~\cite{Julku2016}.

Furthermore, we examined the superfluid weight for the BEC-i and BEC-iii phases within the Popov hydrodynamic theory. This approach not only elucidates the physical origins of different terms in the superfluid weight, but also provides insights into nonperturbative generalizations, such as the renormalization group techniques, and addresses the treatment of infrared divergences inherent in two-dimensional systems with a condensate. By calculating the superfluid weight for various model parameters, we observed that the quantum geometry enhances the superfluid weight by the same order of magnitude as the conventional dispersive term in a narrow band.

Future research directions include conducting a more detailed investigation of the properties of the BEC-ii phase, particularly exploring its effective theory. Additionally, it would be important to develop the low-energy hydrodynamic theories and perturbation theories for the BEC-iii and BEC-i phases. By going beyond the mean-field approximation, we propose employing non-perturbative methods, such as the bosonic dynamical mean-field theory, quantum Monte Carlo or tensor-network approaches, to examine the fate of mean-field phases and assess the validity of the approximations applied in the derivation of the reported results.

\acknowledgements
The authors thank P\"aivi T\"orm\"a,  Sebastiano Peotta, Grazia Salerno, Alexi Julku, Gregor Jotzu, Vincenzo Savona, and Douglas Scalapino for useful discussions. 
I.L. and A.S. acknowledge support from STCU via the IEEE program ``Magnetism for Ukraine 2022'', Grant No. 9918, the National Research Foundation of Ukraine, Grant No.~0120U104963, the Ministry of Education and Science of Ukraine, Research Grant No.~0122U001575, and the National Academy of Sciences of Ukraine, Project No. 0121U108722. A.K. gratefully acknowledges financial support of the Branco Weiss Society in Science, ETH Zurich, through the grant on flat bands, strong interactions, and the SYK physics. 

\bibliography{Refs}

\end{document}